\documentclass[12pt,dvipsnames,a4paper]{article}
\usepackage[top=80pt,bottom=85pt,left=85pt,right=85pt]{geometry}

\pdfoutput=1
\usepackage[utf8]{inputenc}
\usepackage[T1]{fontenc} 
\usepackage[english]{babel} 
\usepackage{braket}
\usepackage[dvipsnames]{xcolor} 
\usepackage{physics}
\usepackage{tabularx} 
\usepackage{ragged2e}
\usepackage{booktabs} 
\usepackage{rotating}
\usepackage{makecell,booktabs}
\usepackage{multirow} 
\usepackage{diagbox}
\usepackage{comment}
\usepackage{mathtools}
\usepackage{amsmath}
\usepackage{cite}
\usepackage{pifont}
\usepackage{bbding}
\usepackage{amssymb}
\usepackage[all]{xy}

\usepackage{relsize}

\usepackage[none]{hyphenat}
\usepackage{tabularray}

\usepackage{tikz}

\usetikzlibrary{math} 
\usetikzlibrary{3d} 
\usetikzlibrary{calc}
\usepackage{tikz-3dplot}
\usepackage{tikz-cd} 
\usetikzlibrary { decorations.pathmorphing, decorations.pathreplacing, decorations.shapes,decorations.markings }\usepackage{microtype}
\usepackage{amsthm}
\usepackage{mathrsfs} 
\usepackage[strict]{changepage}

\tikzcdset{
  cells={font=\everymath\expandafter{\the\everymath\displaystyle}},
}

\newcommand{\xdownarrow}[1]{%
  {\left\downarrow\vbox to #1{}\right.\kern-\nulldelimiterspace}
}

\usepackage{amssymb,amsthm} 
\usepackage{subcaption} 
\DeclareCaptionFormat{custom}

\usepackage[pdftex,colorlinks=true]{hyperref} 
\hypersetup{urlcolor=MidnightBlue, citecolor=burntorange, linkcolor=darkmagenta}

\usepackage{float} 
\usepackage{todonotes}

\numberwithin{equation}{section}

\definecolor{byzantine}{rgb}{0.74, 0.2, 0.64}
\definecolor{burntorange}{rgb}{0.8, 0.39, 0.0}
\definecolor{cambridgeblue}{rgb}{0.64, 0.76, 0.68}
\definecolor{caribbeangreen}{rgb}{0.0, 0.8, 0.6}
\definecolor{celadon}{rgb}{0.67, 0.88, 0.69}
\definecolor{champagne}{rgb}{0.97, 0.91, 0.81}
\definecolor{cream}{rgb}{1.0, 0.99, 0.82}
\definecolor{cyan(process)}{rgb}{0.0, 0.72, 0.92}
\definecolor{brilliantlavender}{rgb}{0.96, 0.73, 1.0}
\definecolor{candypink}{rgb}{0.89, 0.44, 0.48}
\definecolor{darkmagenta}{rgb}{0.55, 0.0, 0.55}

\begin{document}

\begin{titlepage}

\phantom{wowiezowie}

\vspace{-1cm}

\begin{center}

{\Huge {\bf From Multipartite Entanglement }} \\

\vspace{0.3cm}

{\Huge {\bf to TQFT}}

\vspace{1cm}

{\Large Michele Del Zotto$^{\dagger\ddagger*}$, Abhijit Gadde$^{\star}$, and Pavel Putrov$^{\sharp}$ }\\

\vspace{1cm}

{\it
{\small

{\footnotesize$^\dagger$ Mathematics Institute, Uppsala University, \\ Box 480, SE-75106 Uppsala, Sweden}\\
\vspace{.25cm}
{\footnotesize$^\ddagger$ Department of Physics and Astronomy, Uppsala University,\\ Box 516, SE-75120 Uppsala, Sweden}\\
\vspace{.25cm}
{\footnotesize$^{*}$ Center for Geometry and Physics, Uppsala University,\\ Box 516, SE-75120 Uppsala, Sweden}\\
\vspace{.25cm}
{\footnotesize $\star$ Department of Theoretical Physics \\ Tata Institute for Fundamental Research, Mumbai 400005
}\\
\vspace{.25cm}
{\footnotesize $^{\sharp}$ The Abdus Salam International Centre for Theoretical Physics, \\
 Strada Costiera 11, 34151 Trieste, Italy} \\
\vspace{.25cm}
}}

\vskip 1cm
     	{\bf Abstract }
\vskip .1in
\end{center}

At long distances, a gapped phase of matter is described by a topological quantum field theory (TQFT). We conjecture a tight and concrete relationship between the genuine $(d+1)$-partite entanglement --- labelled by a $d$-dimensional manifold $M$ --- in the ground state of a $(d-1)+1$-dimensional gapped theory and the partition function of the low energy TQFT on $M$. 
Under certain assumptions, the conjecture implies that for  $d=3$, the ground state wavefunction can determine the modular tensor category description of the low energy TQFT.
We verify our conjecture for general (2+1)-dimensional Levin-Wen string-net models.

\vfill

\noindent\rule{4cm}{0.4pt}

{\footnotesize \date \\ \noindent Emails: {\tt  michele.delzotto@math.uu.se, abhijit@theory.tifr.res.in, putrov@ictp.it}}

\eject

\end{titlepage}

{
  \hypersetup{linkcolor=black}
  \tableofcontents
}

\section{Introduction and summary}
 In \cite{Kitaev:2005dm,levin2006detecting}, it was shown that a certain linear combination of ground-state entanglement entropies, defined using a decomposition of space into four regions, yields a topological invariant.
 In fact, it computes the three-sphere partition function of the TQFT describing the long-distance physics. This naturally raises the question: is it possible to completely characterize the TQFT from the ground-state wave function? There is mounting evidence that the answer is yes \cite{Zeng:2015pxf, Haah:2014iuw, Kawagoe_2020, Shi:2019mlt, Cian:2022vjb, Jiang_2012, Zhang:2011jd, Li_2008,Liu:2023fuc}. Quantum-information-theoretic tools such as entanglement entropy and modular Hamiltonians - which probe bipartite entanglement - have played a key role in extracting TQFT data from the ground-state wave function. 

There has been recent work \cite{Gadde:2025csh, Gadde:2025ybn, Gadde:2023zni} systematically characterizing and quantifying multi-partite entanglement in quantum states  \cite{Horodecki2009RMP, Amico2008RMP, GuhneToth2009PhysRep, Ma2025FundamentalResearch}. A large class of local-unitary invariants, called multi-invariants, has been effective both for characterizing multipartite entanglement in holographic systems \cite{Gadde:2022cqi, Gadde:2023zzj, Gadde:2024taa} and for extracting genuine multipartite entanglement from them \cite{Iizuka:2025ioc, Iizuka:2025caq}. Interestingly, \cite{Sheffer:2025jtc} observed that a certain family of multi-invariants of the ground state reproduces the partition function of the low-energy TQFT on Lens spaces; see \cite{Sheffer:2025zyr} for an extension to chiral TQFTs. However, the broader problem of completely characterizing the TQFT from the ground-state wave function remains open. In this work, we solidify the relationship between ground-state entanglement and the infrared TQFT.

To state the main result of the paper, we begin with a brief review of the key concepts. Multi-invariants ${\cal Z}(|\psi\rangle)$ are local-unitary invariants of a $q$-partite quantum state $|\psi\rangle$ with the property that they are multiplicative under tensor products: the tensor product of two $q$-partite states defines another $q$-partite state, and ${\cal Z}$ multiplies accordingly. They have a particularly simple construction as polynomials in the wave function and its complex conjugate. They admit a convenient graphical representation as bipartite, $q$-regular, edge-colored graphs, as reviewed in Section~\ref{sec:multi-entropy}. 
This simplicity of construction, together with multiplicativity, makes multi-invariants a useful tool for probing multipartite entanglement. The genuinely $q$-partite component of entanglement - the so-called signal of genuine $q$-partite entanglement, or simply the signal - can be extracted from a multi-invariant by forming certain multiplicative combinations with its “daughters” \cite{AG_signal}. There is some ambiguity in how these combinations are chosen, but one can fix a scheme that is uniquely defined for all multi-invariants. We denote the $q$-partite signal extracted from ${\cal Z}$ using a scheme ${\mathtt S}$ by ${\cal Z}(|\psi\rangle)|_{\mathtt S}$. 

In the paper, we focus on a particular class of multi-invariants that have a nice geometric interpretation. Given a $d$-manifold equipped with a bipartite triangulation\footnote{By bipartite triangulations we mean that the simplices in the triangulation can be colored with two colors such that no two simplices of the same color share a common face.}, we can associate to it a bipartite, $(d+1)$-regular, edge-colored graph. This is done by mapping $d$-simplices to vertices and their codimension-one faces to incident edges, with colors encoding the face labels. Whenever two simplices share a face, the corresponding vertices are connected by an edge of the appropriate color. We review this construction in Section \ref{sec:gem}. 
This graphical encoding of a (bipartitely triangulated) manifold is called a GEM (graph-encoded manifold) \cite{Pezzana1974, Ferri1976, FerriGagliardi1982, FerriGagliardiGrasselli1986, Gagliardi1981RegularEmbeddings, Gagliardi1981RegularGenus, LinsMandel1985, Lins1995, LinsMulazzani2006, BandieriGagliardi2011arXiv}. Since $(d+1)$-partite multi-invariants are likewise specified by graphs of this type, a GEM for a topological manifold $M$ can be used to define a multi-invariant ${\cal Z}(M_\Delta;|\psi\rangle)$, where $\Delta$ denotes a bipartite triangulation of $M$.\footnote{Not all multi-invariants arise from GEMs; see the discussion in Section~\ref{sec:gem}.}

Having described geometric multi-invariants and signals, we now introduce the other main ingredient: the state $|\psi\rangle$. Throughout the paper, we take $|\psi\rangle$ to be $|\psi\rangle_H$, the ground state of a local gapped Hamiltonian $H$ in $(d-1)+1$ dimensions, defined on a $(d-1)$-dimensional space. For example, the $(d-1)$-dimensional space may be a lattice, and $H$ a lattice Hamiltonian.
We further regard $|\psi\rangle_H$ as a $(d+1)$-partite state by partitioning the $(d-1)$-dimensional space into $(d+1)$ regions in a natural way, and treating the Hilbert space supported on each region as a party. This partition is obtained by stereographically projecting the $(d+1)$ faces of a spherical $d$-simplex onto the $(d-1)$-dimensional space. While the precise geometry of the regions will not be important, we take each region to be of size $L$ and work in the large-$L$ limit. The low-energy physics of such a gapped system is expected to be described by a unitary TQFT\footnote{Throughout the paper we will implicitly assume unitarity of the considered TQFTs. It would be interesting to extend our picture to open systems, where the unitarity assumption can be relaxed.}. 
We are now ready to state our main result. Denoting the TQFT partition function on $M$ by\footnote{ We assume that TQFT does not have any framing-type anomalies, that is, the partition function depends only on the diffeomorphism class of the manifold, without any additional structure.} $Z(M)$, we claim the following

\vspace{3ex}
\noindent\textbf{Conjecture:}
\vspace{1ex}
\begin{equation}\label{eq:conj}
    \boxed{
        \lim_{L\rightarrow\infty}{{\cal Z}(M_\Delta;|\psi\rangle_H)|_{\mathtt S}}= \frac{|Z(M)|^2}{Z(S^d)^{n_{\mathtt S, \Delta}}}.
        }
    \end{equation}
    \vspace{2ex}
    
\noindent
Here $Z(S^d)$ is the partition function of the TQFT on a $d$-dimensional sphere $S^d$. Its power $n_{\mathtt S, \Delta}$ depends on the bipartite triangulation $\Delta$ and the scheme $\mathtt S$ used to construct the signal from the multi-invariant, but not on the state $|\psi\rangle_H $. It can be computed explicitly for any $\Delta$ and ${\mathtt S}$. 
In this natural decomposition of the $(d-1)$-dimensional space into $(d+1)$ regions, there is no point common to all regions. Consequently, the genuine $(d+1)$-partite entanglement is necessarily nonlocal. Our conjecture sharpens the slogan that TQFTs capture nonlocal, long-range correlations in the ground state.  

\medskip

In this paper, we provide substantial support for this conjecture by proving that it holds true for all the Levin-Wen string-net models in $d=3$ \cite{Levin:2004mi}, which provide an ideal collection of gapped Hamiltonian lattice models with a known ground state $|\psi\rangle_H$ and a known corresponding TQFT. In this context we can explicitly show a slightly stronger relation
\begin{equation}\label{eq:conj_LV}
    {\cal Z}(M_\Delta;|\psi\rangle_H)|_{\mathtt S} = {|Z(M)|^2 \over Z(S^d)^{n_{\mathtt S, \Delta}}} + O(e^{-cL})
\end{equation}
where $c>0$ is a positive constant we can determine from the data of the underlying spherical fusion category. Moreover, in Appendix \ref{app:alt} we provide a different method for Levin-Wen models that allows to extract the partition function $Z(M)$ itself for all 3-manifolds, not just its absolute value, from the multi-invariant (see formula (\ref{alternative-formula})).

\medskip

Conjecture \eqref{eq:conj} allows us to compute the absolute value of the partition function of the low-energy TQFT on any manifold $M$. In \cite{Sheffer:2026dgj} a method was suggested to extract the argument of $Z(M)$ from multi-invariants itself, assuming $M$ is \textit{locally-achiral}\footnote{Meaning that in it has a graph representation such that all $d$-colored subgraphs are reflection-positive, that is there exist an involution that swaps the vertex colors in the bipartite structure while preserving the colors of the edges. As shown in \cite{Sheffer:2026dgj} not all manifolds have this property when $d=0\mod 4$.}. Together this provides the information about the values $Z(M)\in \mathbb{C}$ of partition functions on closed manifolds. Assuming the values $Z(M)$ are known for \textit{all} closed manifolds, in how much detail does this data determine the TQFT? In Section \ref{sec:univ} we outline how one can possibly recover the full TQFT structure, based on some known facts.


\medskip

The rest of the paper is organized as follows. In Section~\ref{sec:multi-entropy}, we review a class of local-unitary invariants of multipartite quantum states, known as multi-invariants. We will describe how to construct signals of genuine multi-partite entanglement from a general multi-invariant \cite{AG_signal}.
We also explain how to associate a multi-invariant to a triangulated manifold. We review the conditions under which a multi-invariant encodes a manifold \cite{BandieriGagliardi2011arXiv}. In combinatorial topology, this method of describing a manifold is called a graph-encoded manifold (GEM). In Section~\ref{sec:tqft-prelim}  we review the required TQFT preliminaries: the universal construction in Section~\ref{sec:univ} and the Turaev-Viro state-sum model \cite{turaev1992state} in Section~\ref{sec:turaev-viro}. We follow the modern treatment in \cite{kirillov2010turaevviroinvariantsextendedtqft}. In Section \ref{sec:uv_to_ir}, we  compute ${\cal Z}(M_\Delta;|\psi\rangle_H)$  for $4$-region decomposition in a Levin-Wen string-net model. As in many physical examples, treating the ground state as a vector in the factorized Hilbert space associated with four regions requires an extension of the Hilbert space, implemented by relaxing the local Gauss law. We will see this explicitly in the string-net model. 
The resulting multi-invariant yields the partition function of the low-energy TQFT on $M$, together with certain unwanted factors. We show that these extra factors can be understood as arising from bipartite quantum states localized near the boundaries of the regions. We then devise a prescription to remove these contributions, leaving only the contribution from the TQFT partition function on $M$. We comment on several possible generalizations of our approach in Section~\ref{sec:generalizations}. Appendices contain some technical details and examples. 

\section{Multi-invariants and signals}
\label{sec:multi-entropy}
In this section we will quickly review the formalism used to quantify multi-partite entanglement in a general quantum state. Consider a state $|\psi\rangle\in {\cal H}={\cal H}_1\otimes \ldots \otimes {\cal H}_q$. In the context of this paper, $|\psi\rangle$ is going to be the ground state of a microscopic theory e.g. of some lattice model. The lattice is decomposed  into $q$ regions  and each Hilbert space factor ${\cal H}_i$ is the Hilbert space supported on the  $i$-th region. The region decomposition we will be focusing on is the natural decomposition of the $d-1$-dimensional lattice into $d+1$ regions obtained by thinking of the space as the boundary of $d$-dimensional simplex, then the $d+1$ faces of the simplex gives us the $d+1$ regions. This simplex picture of the regions will play an important role when we connect multi-invariants to $d$-manifolds. 

A $q$-partite multi-invariant of $|\psi\rangle$ is a polynomial function ${\cal Z}(|\psi\rangle)$ that is invariant under local unitary transformations i.e. under the action of $U_1\otimes \ldots \otimes U_q$ where $U_i$ is a unitary transformation acting on ${\cal H}_i$, and obey the properties below.
\begin{itemize}
    \item ${\cal Z}(\alpha|\psi\rangle)=|\alpha|^{2n}{\cal Z}(|\psi\rangle)$
    \item ${\cal Z}(|\psi_1\rangle\otimes |\psi_2\rangle)={\cal Z}(|\psi_1\rangle)\cdot {\cal Z}(|\psi_2\rangle)$
\end{itemize}
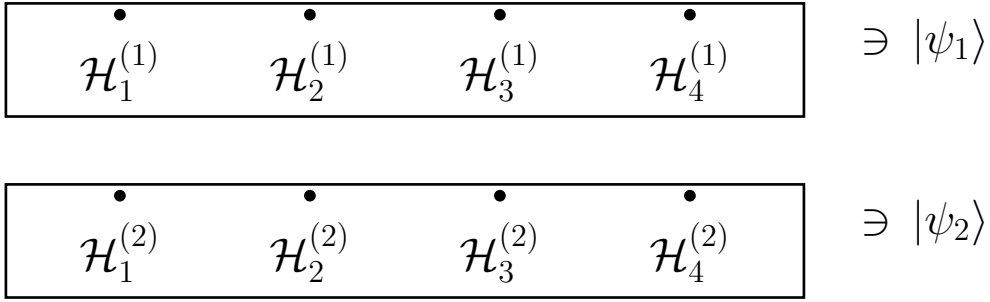
\begin{figure}[t]
  \centering
  \begin{tikzpicture}[every node/.style={font=\Large}]
    \def\W{10.5}   
    \def\H{1.5}    
    \def\sep{2.4}  
    \def\r{2.2pt}  

    \draw[line width=1pt] (0,0) rectangle (\W,\H);
    \draw[line width=1pt] (0,-\sep) rectangle (\W,\H-\sep);

    \foreach \i/\x in {1/1.5,2/4.0,3/6.5,4/9.0}{
      \fill (\x,1.35) circle (\r);
      \node[below=6pt] at (\x,1.35) {${\cal H}_{\i}^{(1)}$};

      \fill (\x,1.35-\sep) circle (\r);
      \node[below=6pt] at (\x,1.35-\sep) {${\cal H}_{\i}^{(2)}$};
    }

    \node[anchor=west] at (\W+0.6,1.0) {$\ni\,|\psi_{1}\rangle$};
    \node[anchor=west] at (\W+0.6,1.0-\sep) {$\ni\,|\psi_{2}\rangle$};

  \end{tikzpicture}
  \caption{Here we have graphically represented the equation $|\psi\rangle=|\psi_1\rangle\otimes |\psi_2\rangle$ where all the states are $q=4$ partite states.}
  \label{fig:parallel_factor}
\end{figure}
In the second equation it is assumed that ${\cal H}={\cal H}^{(1)}\otimes {\cal H}^{(2)}$ and $|\psi_a\rangle\in {\cal H}^{(a)}$ where both ${\cal H}^{(1)}$ and ${\cal H}^{(2)}$ are factorized into $q$ factors. This is shown graphically for $q=4$ in figure \ref{fig:parallel_factor}. The degree of homogeneity $n$ is called the replica number. It is because such a multi-invariant is constructed using $n$ number of copies or ``replicas'' of $|\psi\rangle$ and its dual $\langle\psi|$. To describe the construction in detail, it is convenient to choose a factorized basis and express $|\psi\rangle$ as
\begin{align}
    |\psi\rangle=\sum_{i_1,\ldots, i_q}\psi_{i_1,\ldots, i_q}|i_1\rangle\otimes \ldots \otimes |i_q\rangle
\end{align}
where $i_k$ indexes basis vectors of ${\cal H}_k$. The wavefunction coefficients of the bra are the complex conjugates ${\bar \psi}^{i_1,\ldots, i_q}$. The indices also inform us that $\psi$ transforms as a fundamental representation with respect to all local unitary transformations while $\bar \psi$ transforms, oppositely, as anti-fundamentals. The multi-invariant is constructed by contracting all fundamental indices of any party with the anti-fundamental indices of that party. The resulting quantity is manifestly invariant under local unitary transformations. 

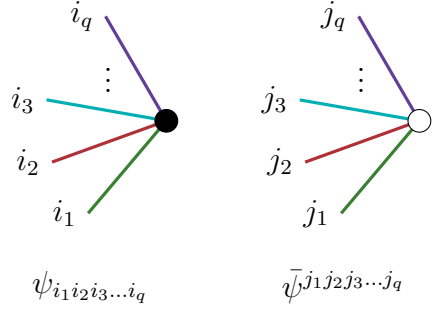
\begin{figure}[t]
  \centering
  \begin{tikzpicture}
        \draw[OliveGreen, very thick] (0,0) -- (-130:1.6) node[left] {\color{black} $i_1$};
        \draw[Maroon, very thick] (0,0) -- (-160:1.6) node[left] {\color{black} $i_2$};
        \draw[TealBlue, very thick] (0,0) -- (-190:1.6) node[left] {\color{black} $i_3$};
        \draw[RoyalPurple, very thick] (0,0) -- (-240:1.6) node[left] {\color{black} $i_q$};
        \draw (-220:1) node  {$\vdots$};
        \draw[fill=black] (0,0) circle (0.15);
        \draw (-1,-2.2) node  {$\psi_{i_1i_2i_3\ldots i_q}$};
  \end{tikzpicture}
  \qquad
   \begin{tikzpicture}
        \draw[OliveGreen, very thick] (0,0) -- (-130:1.6) node[left] {\color{black} $j_1$};
        \draw[Maroon, very thick] (0,0) -- (-160:1.6) node[left] {\color{black} $j_2$};
        \draw[TealBlue, very thick] (0,0) -- (-190:1.6) node[left] {\color{black} $j_3$};
        \draw[RoyalPurple, very thick] (0,0) -- (-240:1.6) node[left] {\color{black} $j_q$};
        \draw (-220:1) node  {$\vdots$};
        \draw[fill=white] (0,0) circle (0.15);
        \draw (-1,-2.2) node  {$\bar{\psi}^{j_1j_2j_3\ldots j_q}$};
  \end{tikzpicture}
  \caption{Graphical presentation of $\psi$ and $\bar \psi$ coefficients. }
  \label{fig:state_graphical}
\end{figure}

\begin{figure}[t]
  \centering
  \begin{tikzpicture}
       \draw[OliveGreen,very thick] (0,0) to (2,0);
       \draw[OliveGreen,very thick] (0,-1.5) to (2,-1.5);
       \draw[OliveGreen,very thick] (0,-3) to[bend left] (2,-3);
       \draw[Maroon,very thick] (0,0) to[bend left] (2,-1.5);
       \draw[Maroon,very thick] (0,-1.5) to (2,0);
       \draw[Maroon,very thick] (0,-3) to[bend right] (2,-3);
       \draw[TealBlue,very thick] (0,0) to[bend right] (2,-1.5);
       \draw[TealBlue,very thick] (0,-1.5) to (2,-3);
       \draw[TealBlue,very thick] (0,-3) to (2,0);
       \draw[fill=white] (2,0) circle (0.15) node[right=0.3] {$1$};
      \draw[fill=white] (2,-1.5) circle (0.15) node[right=0.3] {$2$};
      \draw[fill=white] (2,-3) circle (0.15) node[right=0.3] {$3$};
      \draw[fill=black] (0,0) circle (0.15)  node[left=0.3] {$1$};
      \draw[fill=black] (0,-1.5) circle (0.15) node[left=0.3] {$2$};
      \draw[fill=black] (0,-3) circle (0.15) node[left=0.3] {$3$};
  \end{tikzpicture}
  \caption{A sample $3$-partite multi-invariant with $n=3$.}
  \label{fig:sample}
\end{figure}
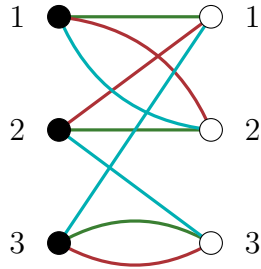

If we label replicas of $\psi$'s as well as of $\bar \psi$'s from $1$ to $n$, then in order to specify the pattern of index contraction of party $1$, we need to specify a permutation $\sigma_1\in S_n$ which indicates that the fundamental index $i_1$ of replica $a\in\{1,\ldots, n\}$ is contracted with the anti-fundamental index of replica $\sigma_1(a)$. Similarly for all other parties. All in all the multi-invariant is described by a permutation tuple $(\sigma_1, \ldots, \sigma_q)$ where all $\sigma_k\in S_n$. The multi-invariant can be written explicitly as
\begin{align}
    {\cal Z}(\sigma_1,\ldots, \sigma_q;|\psi\rangle)=\langle\psi|^{\otimes n} \sigma_1\otimes \ldots\otimes \sigma_q|\psi\rangle^{\otimes n}.
\end{align}
Here we have introduced a new argument for ${\cal Z}$ that is the permutation tuple used to define it.  Note, however, that labeling of multi-invariants by a permutation tuple is ambiguous. This is because the labeling of bra replicas or ket replicas can be changed arbitrarily. We get
\begin{align}
    {\cal Z}(\sigma_1,\ldots, \sigma_q;\cdot)={\cal Z}(g\cdot \sigma_1,\ldots, g\cdot\sigma_q;\cdot)={\cal Z}(\sigma_1\cdot h,\ldots, \sigma_q\cdot h;\cdot),\quad g,h\in S_n.
\end{align}
This ``gauge freedom'' can be fixed by choosing one of the permutations, say $\sigma_1={\rm id}$. 
It is convenient to use a graphical notation to describe a multi-invariant. We denote $\psi$ and $\bar \psi$ by a black and white vertex respectively. Each vertex has $q$ edges of different colors, denoting the $q$ indices, ending on it. This notation is shown in figure \ref{fig:state_graphical}.
If the $i_k$ index of replica $a$ is contracted with the $i_k$ index of replica $b$ then we simply connect $a$-th black vertex with $b$-th white vertex with an edge of color $k$. Finally, we obtain a $q$-regular edge-colored bi-partite\footnote{This ``bi-partite'' is not to be confused with the same word describing entanglement between two parties.} graph describing a multi-invariant. A sample multi-invariant graph is shown in Figure \ref{fig:sample}. 
The permutation tuple for this multi-invariant is $\sigma_{\rm 1}=(1)(2)(3),\;\sigma_{\rm 2}=(12)(3),\;\sigma_{\rm 3}=(123)$. 

Let us also mention that  $1-{\cal Z}^{1/n}$ (where $n$ is the replica number of ${\cal Z}$), for certain multi-invariants termed Coxeter multi-invariants, is monotonically non-increasing under local operations on any party giving it a useful quantum information theoretic interpretation as a good quantifier of multi-partite entanglement \cite{Gadde:2025csh, Gadde:2025ybn}.

\subsection{Genuine multi-partite entanglement signal}
In this section, we will define the genuine multi-partite entanglement and describe the construction of a family of signals that is associated with a particular multi-invariant. We will closely follow the treatment given in  \cite{AG_signal}.

In this section, we will construct a signal of genuine $q$-partite entanglement from multi-invariants. A signal of genuine $q$-partite entanglement, or a $q$-partite signal for short, has the property that it vanishes for any $q$-partite state that is a tensor product of states, each of which is factorized across some bi-partition. For example, consider the following state on parties $A,B$ and $C$,
\begin{align}
    |\psi\rangle_{A,B,C} &=|\psi_1\rangle\otimes |\psi_2\rangle,\notag\\
    |\psi_1\rangle &=(|\psi\rangle_{A_1,B_1}\otimes |\psi\rangle_{C_1}), \, \, |\psi_2\rangle =(|\psi\rangle_{A_2}\otimes |\psi\rangle_{B_2,C_2}).
\end{align}
Here, we have taken each of the Hilbert spaces ${\cal H}_A,{\cal H}_B$ and ${\cal H}_C$ to be factorized as ${\cal H}_A={\cal H}_{A_1}\otimes {\cal H}_{A_2}$ etc. We see that the state $|\psi\rangle_{A,B,C}$ itself is not factorized across parties but $|\psi_1\rangle$ and $|\psi_2\rangle$ both separately are factorized across some bi-partition. In this sense, even though $|\psi\rangle$ itself is not factorized, it does not contain any ``genuine'' $3$-partite entanglement. A $3$-partite signal must vanish on such states. If the signal is additive i.e. if it obeys
\begin{align}
    f(|\psi_1\rangle\otimes |\psi_2\rangle)=f(|\psi_1\rangle)+f(|\psi_2\rangle),
\end{align}
then we only need to ensure that it vanishes on factorized states. In fact, we only have to impose that it vanishes if $q$-partite state is factorized as $q-1$-partite state and $1$-partite state. To understand why this is the case, consider a $4$-partite state factorized as $|\psi\rangle_{A,B,C,D}=|\psi\rangle_{A,B}\otimes |\psi\rangle_{C,D}$. The right hand side can also be written as,
\begin{align}
    |\psi\rangle_{A,B,C,D}=(|\psi\rangle_{A_1,B_1}\otimes|0\rangle_{C_1}\otimes |0\rangle_{D_1}) )\otimes(|0\rangle_{A_2}\otimes |0\rangle_{B_2}\otimes |\psi\rangle_{C_2,D_2}).
\end{align}
Here we have the original parties $A,B,C,D$ have been renamed as $A_1,B_1, C_2, D_2$ respectively and extra $1$-dimensional factors $|0\rangle_{A_2}, |0\rangle_{B_2}, |0\rangle_{C_1}, |0\rangle_{D_1}$ in each party have been introduced. From the new expression for $|\psi\rangle_{A,B,C,D}$, it is clear that the $4$-partite additive signal vanishes on it, because it vanishes on any state that is factorized as a $3$-partite state and a $1$-partite state.

If a $q$-partite state is a tensor product of states one of which is not factorized and the rest are factorized, then an additive signal extracts the contribution from the piece that is not factorized. This makes an additive $q$-partite signal a good diagnostic of the multi-partite entanglement structure of the state. The ground state of wavefunction of a $d$-dimensional (including time) many body system thought of as a $d+1$-partite state with the parties being the faces of a $d$-simplex, can be understood as effectively containing a ``non-local'' or ``long-range'' piece that is purely $d+1$-partite and the rest of the lower partite pieces localized at loci where multiple regions meet. An additive signal can then be used to extract the contribution of the non-local, non-factorized piece.

\subsection{Construction of additive signals}\label{construction_signal}
As a multi-invariant is multiplicative, its logarithm is additive. We introduce a separate notation for it,
\begin{align}
    {\cal E}(\sigma_1,\ldots, \sigma_q;|\psi\rangle):=-\log  {\cal Z}(\sigma_1,\ldots, \sigma_q;|\psi\rangle). 
\end{align}
We will construct $q$-partite signals as linear combinations of logarithmic multi-invariants. As we emphasized earlier, we simply need to impose that it vanishes for $q$-partite states that factorize as a $q-1$-partite state and a $1$-partite state. This needs to be true for the factorization of each of the $q$-parties. A multi-invariant whose permutations are chosen from the set $\Sigma=\{\sigma_1,\ldots, \sigma_q\}$ of $q$ permutations is called multi-invariant of family $\Sigma$. 
In the linear combination, we will only consider logarithmic multi-invariants of the same family.

First note that, for any choice of $\Sigma$,
\begin{align}\label{factorize_vanishing}
    {\cal E}(\sigma_i,\sigma_2,\ldots, \sigma_q;\cdot)-{\cal E}(\sigma_j,\sigma_2,\ldots, \sigma_q;\cdot)=0, 
\end{align}
evaluated on the state where the party $A_1$ is factorized from the rest.\footnote{This equation is true even if $\sigma_i$ and $\sigma_j$ are $S_n$ permutations not chosen from $\Sigma$.} This equation motivates us to consider a formal linear combination of permutations defined as,
\begin{align}
    {\cal E}(\sum_i a_i\sigma_i, \ldots;\cdot)= \sum_i a_i{\cal E}(\sigma_i,\ldots;\cdot).
\end{align}
This definition is extended to all the other permutation arguments of ${\cal E}$. In this notation, equation \eqref{factorize_vanishing} becomes
\begin{align}
    {\cal E}(\sum_i a_i\sigma_i, \ldots;\cdot)=0,\qquad{\rm with}\, \,  \sum_i a_i=0,
\end{align}
where the rest of the arguments, denoted by $\ldots$, can be any linear combination of permutations. It immediately follows that
\begin{align}
    {\cal E}(\sum_i a_i^{(1)} \sigma_i, \sum_i a_i^{(2)} \sigma_i, \ldots, \sum_i a_i^{(q)} \sigma_i;\cdot), \qquad {\rm with}\,\,  \sum_i a_i^{(k)}=0, \forall k
    \label{signal-general}
\end{align}
is an additive signal. By virtue of linearity, a linear combination of the above additive signals is also an additive signal. This also makes it clear that there is a large amount of freedom in constructing the signal and our formula parametrizes that freedom explicitly. A particularly simple form is constructed by taking differences of two permutations in each argument. This gives $2^q$ terms in the linear combination. For example, for $4$-parties, a simple signal is
\begin{align}
    {\cal E}(\sigma_1-\sigma_2, \sigma_2- \sigma_1, \sigma_3-\sigma_1,  \sigma_{4}-\sigma_1;\cdot).
\end{align}
We will use this signal to extract topological partition in Section \ref{sec:uv_to_ir}.

\subsection{Connection to triangulated manifolds}\label{sec:gem}
Consider a closed oriented $d$-dimensional topological manifold that is triangulated with $d$-simplexes. Let us also assume that the triangulation is bi-partite. For any piecewise-linear manifold, it is possible to pick such a triangulation. Because of the bi-partite-ness, we can color the simplex black and white alternatingly, so that any white-colored simplex is an orientation-reversed copy of any black-colored simplex and vice versa. A multi-invariant can be assigned to such a triangulation quite straightforwardly. We simply take a white (black) simplex to be a white (black) vertex with each of the faces representing edges of different colors ending on it. In this way, the bi-partite triangulation maps to a $q$-regular, edge-colored, bi-partite graph representing a multi-invariant. This combinatorial description of a manifold is known as a ``graphically encoded manifold'' or GEM for short. These GEMs have been widely studied in the past. In the context of multi-invariants, a vertex of the graph is 
the $q$-partite quantum state on $q-1$-dimensional  simplex with each party being associated to one of its $q$ faces. The number of white (or equivalently black) simplexes is the replica number of multi-invariant. 

What about the converse question, can we associate a triangulated $q-1$-dimensional manifold to any $q$-partite multi-invariant? The answer to this question is more subtle. For $q=3$, indeed all multi-invariants can be associated to $q-1=2$-manifolds, but for $q\geq 4$ multi-invariants need to be obey certain conditions to be associated to $q-1$-dimensional manifolds. Let us understand this by taking the example of $q=4$. In this case, the state is given by a $3$-simplex, each of its four faces representing a party. The question is whether the gluing of faces as prescribed by the multi-invariant describes a non-singular manifold. In order to see this, let us cut all corners of the $3$-simplex. After gluing all the truncated simplexes, we will get a $3$-manifolds with some holes coming from the cut corners. As we fill out the corner spaces by growing the corners, we find ourselves with an obstruction if the boundary of any hole is not a sphere, because the link of a point in the manifold has to be a sphere. This obstruction also indicates the solution. As the manifold with holes is triangulated, the boundary of the hole is also triangulated. This boundary is what one obtains if one takes a particular $3$-colored subgraph of the multi-invariant graph and interprets it as a multi-invariant of a $3$-partite state. For boundary of all the holes to be spheres we impose that the $2$-manifold associated to any $3$-colored subgraph must be of spherical topology. In this way, one can characterize the conditions on the multi-invariant recursively. Note that for $q=3$, there is no obstruction because all boundaries of the holes are circular and can be filled in canonically. In fact, for a general $q$, the condition that all $q-1$-colored subgraphs must correspond to $q-2$ dimensional spheres is sufficient for the $q$-partite multi-invariant to be describing a $q-1$-dimensional manifold \cite{BandieriGagliardi2011arXiv}. Multi-invariant graphs that satisfy this condition are also called ``graph encoded manifolds'' or GEMs for short. They offer a combinatorial way to study topology. If a multi-invariant ${\cal Z}(\sigma_1,\ldots, \sigma_q;\cdot)$ is a GEM, then ${\cal Z}(\sigma_1',\ldots, \sigma_q';\cdot)$ where $\sigma'$s are a permutation of $\sigma$s is also a GEM and represents the same triangulated manifold or its orientation reversal, depending on the parity of the permutation. This is because all faces of a $d$-simplex are equivalent, and their order determines its orientation. The manifold depends only on the collection $\Sigma=(\sigma_1, \ldots, \sigma_q)$ up to an even permutation. The relation between various notions on the multi-invariant graph and triangulated manifold is summarized in the Table \ref{tab:dual_graph}. Many explicit examples of GEMs realizing particular 2- and 3-manifolds can be found in the book \cite{Lins1995}. Some of them are listed in Appendix \ref{app:more-examples}.
\begin{table}[t]
  \centering
  \begin{tabular}{l|l}  
    \hline
    Triangulations & Multi-invariant graph \\
    \hline
    $3$-simplex & Vertex \\
    Face & $1$-colored subgraph (edge) \\
    Edge &  $2$-colored subgraph (loop) \\
    Vertex & $3$-colored subgraph\\
    \hline
  \end{tabular}
  \caption{Relation between bi-partitely triangulated manifold and multi-invariant graph.}
  \label{tab:dual_graph}
\end{table}

In order to compute the partition function of the TQFT on a manifold $M$, we will consider a multi-invariant ${\cal Z}$ that is a GEM of $M$. Of course, such a GEM is not unique, and the multi-invariant depends on the choice of the bi-partite triangulation and not just on the topology of $M$. 
However, for three-dimensional non-chiral theories, we will show that by linearly combining ${\cal E}=-\log {\cal Z}$'s of the same family $\Sigma$, we get a quantity that depends only on the topology of $M$. Any genuine multi-partite signal constructed from ${\cal E}$ from the $\Sigma$ family has this property.

\section{TQFT preliminaries}
\label{sec:tqft-prelim}

In this section we review some basics about TQFTs. Section \ref{sec:univ} is about  recovering TQFT from the knowledge of the partition functions on closed $d$-manifolds, while Section \ref{sec:turaev-viro} contains a quick review of Turaev-Viro TQFTs.

\subsection{Universal construction}
\label{sec:univ}

The non-extended TQFT structure (as defined by Atiyah-Segal axioms), under some assumptions, is known to be recoverable from the knowledge of partition functions on closed manifolds by means of the \textit{universal construction}, originally introduced in \cite{blanchet1995topological} (see also \cite{turaev2016quantum,khovanov2020universal}). The basic idea is as follows.  As the first step defines an \textit{unphysical} Hilbert space associated to a closed $(d-1)$-manifold $\Sigma$ as freely generated by $d$-manifolds with boundary $\Sigma$:
\begin{equation}
    \mathcal{H}_\text{unphys}(\Sigma):=\bigoplus_{\partial M=\Sigma} \mathbb{C}|M\rangle.
\end{equation}
The knowledge of the partition function on closed manifolds provides it with a natural pairing:
\begin{equation}
    \langle M'|M\rangle :=Z(\bar{M}'\cup_{\Sigma}M)
\end{equation}
where the bar denotes the orientation reversal. Next, we take its quotient over the kernel of the pairing:
\begin{equation}
    \mathcal{H}(\Sigma):=\mathcal{H}_\text{unphys}(\Sigma)/\sim,\qquad |\psi\rangle \sim |\psi'\rangle\Leftrightarrow \langle \phi|\psi\rangle =\langle \phi|\psi'\rangle,\;\forall |\phi\rangle.
    \label{hilb-quot}
\end{equation}
If the original TQFT is \textit{non-degenerate} (also known as \textit{cobordism generated}), that is its Hilbert space on $\Sigma$ is generated by the states produced by manifolds with boundary $\Sigma$, then $\mathcal{H}(\Sigma)$ recovers that physical Hilbert space.

The evolution map associated to a $d$-dimensional bordism $M$ from $\Sigma$ to $\Sigma'$ (i.e. $\partial M=\bar\Sigma\sqcup \Sigma'$) is then provided by
\begin{equation}
    \begin{array}{rrcl}
         Z(M):& \mathcal{H}(\Sigma)&\longrightarrow & \mathcal{H}(\Sigma'),  \\
         & \big[\sum_ic_i|M_i\rangle\big]& \longmapsto & \big[\sum_ic_i|M\cup _{\Sigma}M_i\rangle\big],
    \end{array}
\end{equation}
where $\partial M_i=\Sigma$ and therefore $\partial (M\cup _{\Sigma}M_i) =\Sigma'$. We also remark that just the dimensions of the Hilbert spaces (i.e. ground state degeneracies) can be simply recovered by the well-known relation
\begin{equation}
    \dim\mathcal{H}(\Sigma)=Z(\Sigma\times S^1).
\end{equation}
Then, in practice, one can choose a basis in $\mathcal{H}(\Sigma)$ to be represented by a collection of $\dim\mathcal{H}(\Sigma)$ generic enough 3-manifolds with boundary $\Sigma$.

It can (and often does) happen however that the TQFT is \textit{degenerate}. Meaning there are states that cannot be realized by any manifolds \textit{without insertion of extended operators}. This can be a consequence of a non-trivial global symmetry that forbids creation of charged (i.e. transforming non-trivially under the symmetry) states in this way. In the case of $d=3$ in particular, this will happen already for $\Sigma=T^2$ if there is a symmetry exchanging simple line operators. This is because empty manifolds can only produce states that are invariant under the symmetry, which is, however, not true for manifolds (e.g. solid tori) with line operators inserted. Mathematically, such a symmetry is an autoequivalence of the modular tensor category (MTC) that describes the line operators\footnote{An example of an anomaly free MTC which has such an autoequivalence is the Drinfeld center of $\mathrm{Vec}_{\mathbb{Z}_2}$ -- that is untwisted $\mathbb{Z}_2$ gauge theory. There is an order two autoequivalence that exchanges the Wilson and 't Hooft lines. An example that does not have such an autoequivalence is the Fibbonaci category or its Drinfeld center.}. In this case, the vector space (\ref{hilb-quot}) produced by the universal construction instead satisfies\footnote{Note that as an input of the universal construction one can in principle take an arbitrary numerical invariant of $d$-manifolds which is multiplicative with respect to the disjoint union operation. However, if this invariant is not a partition function of a non-degenerate TQFT, the output is what is sometimes referred to as \textit{lax-monoidal} TQFT or \textit{topological theory} which fails some of the Atiyah-Segal axioms. In particular, the Hilbert space associated with the disjoint union of $(d-1)$-manifolds is not necessarily the tensor product of the Hilbert spaces associated with the connected components.}
\begin{equation}
    \dim\mathcal{H}(\Sigma)<Z(\Sigma\times S^1).
\end{equation}
and an additional step is needed to recover the full physical Hilbert space\cite{McNamara:2026isz}. The crucial part is considering the spaces $\mathcal{H}(\Sigma)$ from the universal construction for $\Sigma$ being a disjoint union of $(d-1)$-manifolds, which allows one to produce states on the individual components that are not constrained by superselection rules.

This method in principle can be applied on higher levels to recover extended TQFT structure. In particular, in the case of $d=3$ one can use the extended universal construction described in\footnote{In our setting, the input numerical invariant would be defined on \textit{nondecorated} closed 3-manifolds.} \cite{de2017construction} to obtain the 3-2-1 TQFT structure (i.e. 3d TQFT extended down to 1-manifolds) provided by a 2-functor on the 2-category of 3-2-1-bordisms. Below we outline a somewhat different approach to recovering the modular tensor category (MTC) that governs the 3-2-1 extended TQFT structure, assuming one already has the knowledge of the non-extended TQFT. 

Physically, the MTC can be understood as the category of the line operators in the 3d TQFT\footnote{This category is not to be confused with the (generically not modular, or even braided) spherical fusion category $\mathcal{C}$ which will appear later in the paper as the input of the Turaev-Viro/Levin-Wen construction. The category of the line operators in the 3d TQFT obtained by such a construction is the \textit{Drinfeld center} of $\mathcal{C}$.}. The first step is the recovery of the set of simple objects (see Section \ref{sec:turaev-viro} for a basic review of the categorical description of topological line operators). They correspond to a special basis in $\mathcal{H}(T^2)$, the Hilbert space on a torus. This basis can be specified as follows. Consider the (non-extended) 2d TQFT obtained by compactifying the non-extended 3d TQFT. Its Hilbert space on $S^1$ is $\mathcal{H}(T^2)$. An arbitrary unitary 2d TQFT is known to be a direct sum of invertible 2d TQFTs. The basis that provides such a decomposition is the special basis in $\mathcal{H}(T^2)$ corresponding to simple line operators in the 3d TQFT. Namely, the special basis states on the torus are realized by the solid torus with a simple line operator inserted at its core. The contractible loop (``meridian'') on the solid torus boundary is identified with the compactification circle. After the set of simple objects is understood in this way, the knowledge of the non-extended TQFT provides us with the knowledge of the partition functions on closed 3-manifolds with collections of simple line operators inserted. Taking now this as an input of the universal construction, one obtains the TQFT defined as a functor on the 1-category of bordisms decorated with lines labeled by simple objects. In particular, it provides a Hilbert space on Riemann surfaces with points marked by simple objects. The fusion spaces of line operators are then given by Hilbert spaces on a sphere with marked points. After this, one has the knowledge of what TQFT associates with bordisms decorated by arbitrary graphs, with edges labeled by simple line operators and junctions labeled by elements of the fusion spaces. This is achieved by removing a small ball centered at each junction and putting on its boundary a corresponding state in the Hilbert space of the sphere with marked points. Such a knowledge, in particular, allows one to recover the $F$- and $R$-symbols of the MTC, using one of their standard definitions\footnote{The $F$-symbol, for example, can be recovered as the expectation value of the graph shown in Figure \ref{fig:6j} (b).}.

\subsection{Turaev-Viro TQFT}
\label{sec:turaev-viro}

The original Turaev-Viro TQFT construction \cite{turaev1992state} used quantum $6j$ symbols for $\mathfrak{sl}_2$ and the partition function on a 3-manifold was defined in terms of its triangulation. It was later generalized to a construction with the input being an arbitrary spherical fusion category $\mathcal{C}$ \cite{barrett1996invariants}. The original construction is then recovered when $
\mathcal{C}$ is (a semisimplification of) the category of finite-dimensional representations of the quantum group $\bar{U}_q(\mathfrak{sl}_2)$ for a root of unity $q$. In \cite{kirillov2010turaevviroinvariantsextendedtqft} a construction of the extended Turaev-Viro TQFT via \textit{polytope decompositions} (which generalize triangulations) was given. We will be using this latter approach as it is more naturally can be related to the string-net model of Levin-Wen \cite{kirillov2011stringnetmodelturaevviroinvariants}. Below we review the basics needed in the context of this paper and refer to \cite{kirillov2010turaevviroinvariantsextendedtqft} for all the details. 

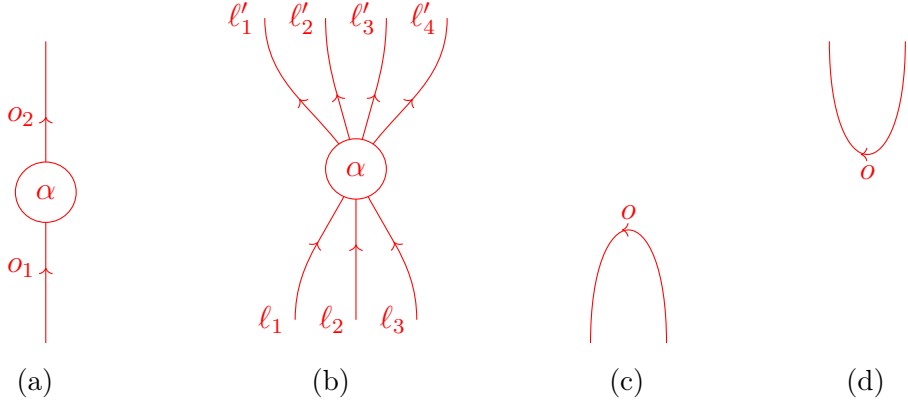
\begin{figure}[ht!]
    \centering
  \begin{subfigure}{0.2\textwidth}
        \centering
        \begin{tikzpicture}
          \begin{scope}[red,decoration={markings,mark=at position 0.5 with {\arrow{>}}}]
        \draw[postaction={decorate}] (0,-2) -- (0,0) node[midway,left] {$o_1$};
        \draw[postaction={decorate}] (0,0) -- (0,2)  node[midway,left] {$o_2$};
       
    \end{scope}
    \filldraw[color=red,fill=white] (0,0) circle (0.4) node {${\alpha}$};
        \end{tikzpicture}
        \caption{}
    \end{subfigure}
    \begin{subfigure}{0.3\textwidth}
        \centering
        \begin{tikzpicture}
           \begin{scope}[red,decoration={markings,mark=at position 0.5 with {\arrow{>}}}]
        \draw[postaction={decorate}] (-0.8,-2) node[left]{$\ell_1$} to[out=90,in=-110] (0,0) ;
         \draw[postaction={decorate}] (0,-2) node[left]{$\ell_2$} -- (0,0);
         \draw[postaction={decorate}] (0.8,-2) node[left]{$\ell_3$} to[out=90,in=-70] (0,0) ;

     \draw[postaction={decorate}] (0,0)  to[out=120,in=-90] (-1.2,2) node[left]{$\ell_1'$};
     \draw[postaction={decorate}] (0,0)  to[out=100,in=-90] (-0.4,2) node[left]{$\ell_2'$};
     \draw[postaction={decorate}] (0,0)  to[out=80,in=-90] (0.4,2) node[left]{$\ell_3'$};
     \draw[postaction={decorate}] (0,0)  to[out=60,in=-90] (1.2,2) node[left]{$\ell_4'$};
       
    \end{scope}
    \filldraw[color=red,fill=white] (0,0) circle (0.4) node {${\alpha}$};
        \end{tikzpicture}
        \caption{}
    \end{subfigure}
  \begin{subfigure}{0.2\textwidth}
        \centering
        \begin{tikzpicture}
          \begin{scope}[red,decoration={markings,mark=at position 0.5 with {\arrow{<}}}]
        \draw[postaction={decorate}] (-0.5,-2) .. controls (-0.5,0) and (0.5,0) .. (0.5,-2) node[midway,above] {$o$};

    \end{scope}
    
        \end{tikzpicture}
        \caption{}
    \end{subfigure}
    \begin{subfigure}{0.2\textwidth}
        \centering
        \begin{tikzpicture}
          \begin{scope}[red,decoration={markings,mark=at position 0.5 with {\arrow{<}}}]
        \draw[postaction={decorate}] (-0.5,2) .. controls (-0.5,0) and (0.5,0) .. (0.5,2) node[midway,below] {$o$};
        \draw (0,-2);

    \end{scope}
    
        \end{tikzpicture}
        \caption{}
    \end{subfigure}
  
    \caption{(a) The graphical representation of $\alpha\in \mathrm{Hom}_\mathcal{C}(o_1,o_2)$. (b) An equivalent diagram in the case when $o_1=\ell_1\otimes \ell_2\otimes \ell_3$ and $o_2=\ell_1'\otimes \ell_2'\otimes \ell_3'\otimes \ell_4'$. (c) The graphical representation of the evaluation morphism $o^*\otimes o\rightarrow\mathbf{1}$. (d) The graphical representation of the coevaluation morphism $\mathbf{1} \rightarrow o\otimes o^*$.}
    \label{fig:hom}
\end{figure}

We begin with an informal review of the notion of spherical fusion category. It can be understood as the mathematical structure that describes topological line operators in 2 dimensions. The objects of the category correspond to the line operators themselves. The morphisms $\mathrm{Hom}_\mathcal{C}(o_1,o_2)$ between two line operators $o_1$ and $o_2$ are the space of point-like operators that can be supported on the junction between $o_1$ and $o_2$ (see Figure \ref{fig:hom} (a)). There is distinguished finite collection  of \textit{simple} line operators $\ell\in  \mathrm{Irr}(\mathcal{C})$ such that an arbitrary line operator decomposes as
\begin{equation}
    o=\sum_{\ell\in\mathrm{Irr}(\mathcal{C})}n^o_\ell\,\ell,\qquad
    n^o_\ell\in \mathbb{Z}_{\geq 0}.
\end{equation}
Moreover, there are no morphisms between two different simple objects: $\mathrm{Hom}(\ell,\ell')=\emptyset$, $\ell\neq\ell'$, and the only morphisms between a simple object and itself are multiples of the identity morphism: $\mathrm{Hom}(\ell,\ell)=\mathbb{C}\,\mathrm{Id}_\ell$. The category also has a monoidal operation $\otimes:\mathcal{C}\times \mathcal{C}\rightarrow \mathcal{C}$ which corresponds to fusing parallel line operators (see Figure \ref{fig:hom} (b)). The monoidal identity $\mathbf{1}$ is the simple object corresponding to a trivial line operator. There is also a \textit{duality} operation $(\;\;)^*:\mathcal{C}\rightarrow \mathcal{C}$ that corresponds to reversing the orientation of the line operators. There are distinguished \textit{evaluation} $o^*\otimes o\rightarrow \mathbf{1}$ and \textit{coevaluation} and $\mathbf{1}\rightarrow o\otimes o^*$ realized by ``folding'' the line operator $o$ (see Figure \ref{fig:hom}). Their composition, corresponding to the insertion of $o$ supported on a contractible circle, gives the morphism $d_{o}\,\mathrm{Id}_o\in \mathrm{Hom}_\mathcal{C}(\mathbf{1},\mathbf{1})$, where $d_{o}\in \mathbb{C}$ is the \textit{quantum dimension} of the line operator $o$. Unitarity requires $d_{o}\in \mathbb{R}_{>0}$. All the above operations satisfy the conditions necessary to self-consistently describe line operators on arbitrary 2d spacetime. In particular $d_o=d_{o^*}$, since, on a sphere, $o$ supported on a circle can also be viewed as $o^*$ supported on a circle with the same orientation.

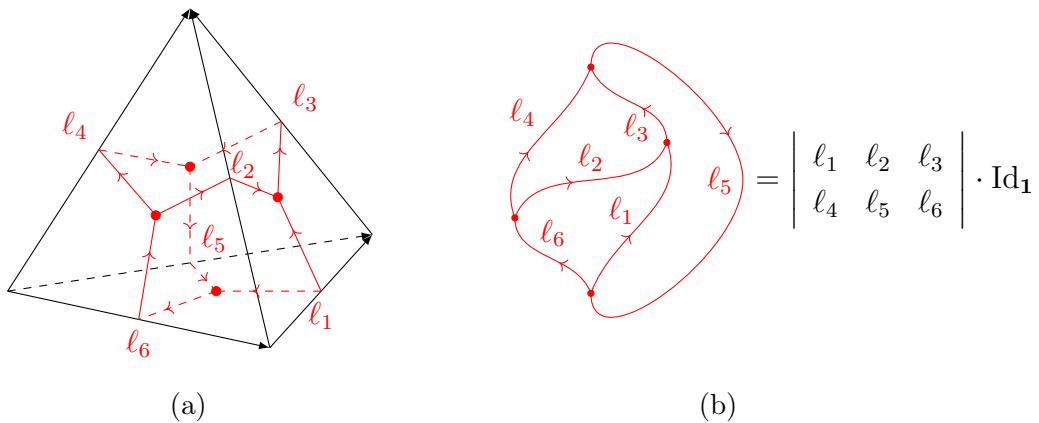
\begin{figure}[ht!]
    \centering
 
   \begin{subfigure}{0.4\textwidth}
        \centering
        \begin{tikzpicture}[scale=1.5]
             \draw[-latex] (0,0) -- (2.3,-0.5);
            \draw[-latex] (2.3,-0.5) -- (3.2,0.5);
            \draw[-latex] (3.2,0.5) -- (1.6,2.5);
            \draw[-latex] (0,0) -- (1.6,2.5);
            \draw[-latex] (2.3,-0.5) -- (1.6,2.5);
            \draw[-latex,dashed] (0,0) -- (3.2,0.5);
    \begin{scope}[red,decoration={markings,mark=at position 0.65 with {\arrow{>}}}]
    \draw[postaction={decorate}] (1.3,0.67) -- (0.8,1.25) node[above left]{$\ell_4$};
    \draw[postaction={decorate}] (1.15,-0.25) node[below]{$\ell_6$} -- (1.3,0.67);

    \draw[postaction={decorate}] (2.75,0) node[below]{$\ell_1$} -- (2.37,0.83);

    \draw[postaction={decorate}] (2.37,0.83) -- (2.4,1.5) node[above right]{$\ell_3$};

    \draw[postaction={decorate}] (1.95,1) node[above right=-0.2]{$\ell_2$} -- (2.37,0.83);

   \draw[postaction={decorate}] (1.3,0.67) -- (1.95,1);

  \draw[postaction={decorate},dashed] (0.8,1.25) -- (1.6,1.1);

    \draw[postaction={decorate},dashed] (2.4,1.5) -- (1.6,1.1);

    \draw[postaction={decorate},dashed] (1.6,1.1) -- (1.6,0.25);

    \draw[postaction={decorate},dashed] (1.6,0.25) node[above right]{$\ell_5$} -- (1.83,0);

    \draw[postaction={decorate},dashed] (1.83,0) -- (1.15,-0.25);

    \draw[postaction={decorate},dashed] (2.75,0) -- (1.83,0);
    
    \end{scope}
    \draw[red,fill=red] (1.6,1.1) circle (0.04);
    \draw[red,fill=red] (1.3,0.67) circle (0.04);
    \draw[red,fill=red] (2.37,0.83) circle (0.04);
    \draw[red,fill=red] (1.83,0) circle (0.04);
        \end{tikzpicture}
        \caption{}
    \end{subfigure}
    \qquad
    \begin{subfigure}{0.4\textwidth}
        \centering
        \begin{tikzpicture}
        \begin{scope}[red,decoration={markings,mark=at position 0.4 with {\arrow{>}}}]
            \draw[postaction=decorate] (-1,-0.5) to[out=70,in=-110] (1,0.5);
            \draw[postaction=decorate] (-1,-0.5) to[out=110,in=-110] (0,1.5);
            \draw[postaction=decorate] (0,-1.5) to[out=110,in=-90] (-1,-0.5);
            \draw[postaction=decorate] (0,-1.5) to[out=70,in=-70] (1,0.5);
            \draw[postaction=decorate] (1,0.5) to[out=90,in=-70] (0,1.5);
            \draw[postaction=decorate] (0,1.5) to[out=90,in=90] (2,0) to[out=-90,in=-90]  (0,-1.5);

            \draw (-0.9,0.9) node {$\ell_4$};
            \draw (0,0.3) node {$\ell_2$};
            \draw (0.6,0.7) node {$\ell_3$};
            \draw (-0.5,-0.7) node {$\ell_6$};
            \draw (0.4,-0.4) node {$\ell_1$};
            \draw (1.7,0) node {$\ell_5$};
            
        \end{scope}
        \draw (4,0) node {$=\left|\begin{array}{ccc}
            \ell_1 & \ell_2 & \ell_3  \\
             \ell_4 & \ell_5 & \ell_6
        \end{array}\right|\cdot \mathrm{Id}_{\mathbf{1}}$};
        \draw[red,fill=red] (-1,-0.5) circle (0.04);
        \draw[red,fill=red] (1,0.5) circle (0.04);
        \draw[red,fill=red] (0,-1.5) circle (0.04);
        \draw[red,fill=red] (0,1.5) circle (0.04);
        \end{tikzpicture}
        \caption{}
    \end{subfigure}
  
    \caption{(a) The dual graph $\mathcal{G}$ for a tetrahedron show in red. The edges are decorated by simple objects $\ell_i\in \mathrm{Irr}(\mathcal{C})$ and vertices by the morphisms from $\mathrm{Hom}_{\mathcal{C}}(\ell_i\otimes \ell_j,\ell_k)$ or $\mathrm{Hom}_{\mathcal{C}}(\ell_k,\ell_i\otimes \ell_j)$. (b) The graph $\mathcal{G}$ embedded in a plane. Read from bottom to top, it provides a composition of the morphisms of the form shown in Figure \ref{fig:hom}. The result is a morphism $\mathbf{1}\rightarrow \mathbf{1}$, which is necessarily a multiple of the identity. In the case when $\mathcal{C}$ is (the semisimplification of) the category of finite-dimensional representations of the quantum $\mathfrak{sl}_2$ at a root of unity, the labels $\ell_i$ can be identified with half-integer spins and the choice of morphisms at the trivalent junctions is unique (up to a multiple). With the standard choice (given by quantum Clebsch-Gordon coefficients), the result of the composition is then given by a properly normalized quantum \textit{6j}-symbol \cite{turaev1992state}.}
    \label{fig:6j}
\end{figure}

The basic building block of the Turaev-Viro TQFT is the value $Z_{TV}(P,\ell,\alpha)\in \mathbb{C}$ assigned to an oriented polyhedron $P$ with certain ``colorings'' $\ell$ and $\alpha$ of its edges and faces, respectively. Namely, to each oriented edge $e\in \mathrm{Edges}(P)$ we assign a simple object $\ell(e)\in \mathrm{Irr}(\mathcal{C})$. The edge with the reversed orientation is assigned with the dual object: $\ell(\bar{e})=\ell(e)^*$. Each face $f\in \mathrm{Faces}(P)$ possesses the orientation induced from the orientation of $P$. To a face $f$ we then assign a basis element $\alpha_f\in \mathrm{Hom}_\mathcal{C}(\ell(e_1)\otimes \ldots \otimes \ell(e_N),\mathbf{1})$ where $e_i\in \partial f$ are the edges on the boundary of $f$ with the induced orientation. The value $Z_{TV}(P,\ell,\alpha)$ is then defined as follows. Consider the planar graph $\mathcal{G}$ dual to the polygonal decomposition of the $\partial P$ with the edges of the dual graph oriented arbitrarily. The coloring $\ell$ then provides the coloring of the edges of the dual graph by simple objects of $\mathcal{C}$. Namely, an edge of $\mathcal{G}$ is colored by $\ell(e)$ where $e$ is the edge of the polyhedron $P$ intersected by the considered edge of $\mathcal{G}$, with the consistent orientation. See Figure \ref{fig:6j} (a) for the dual graph for a tetrahedron shown in red. To each vertex of $\mathcal{G}$ we then assign the morphism $\alpha_f$ associated with the dual face. The planar graph $\mathcal{G}$ with such colorings then provides a morphism in $\mathrm{Hom}_\mathcal{C}(\mathbf{1},\mathbf{1})= \mathbb{C}\,\mathrm{id}_{\mathbf{1}}$ via the graphical calculus for spherical fusion categories briefly reviewed above (with the result independent of various choices made), see Figure \ref{fig:6j} for an example. The value of $Z_{TV}(P,\ell,\alpha)\in \mathbb{C}$ is defined to be the coefficient of proportionality between the morphism and $\mathrm{id}_\mathbf{1}$. 

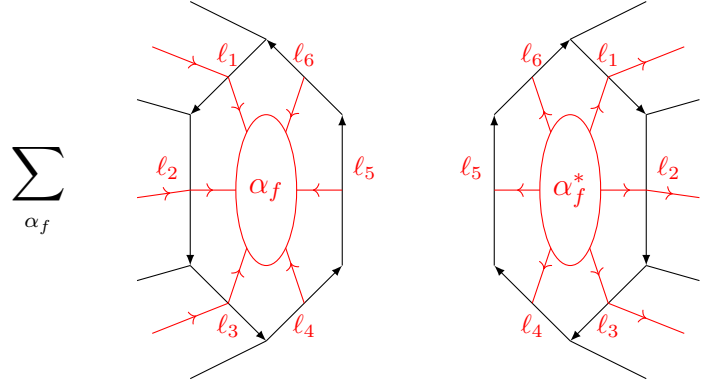
\begin{figure}[ht!]
    \centering
\begin{tikzpicture}
    \draw (-5,0) node {$\mathlarger{\mathlarger{\sum}}\limits_{\alpha_f}$};
    \draw[-latex] (-1,-1) --  (-1,1);
    \draw[-latex] (-1,1) -- (-2,2);
    \draw[-latex] (-2,2) -- (-3,1);
    \draw[-latex] (-3,1) -- (-3,-1);
    \draw[-latex] (-3,-1) -- (-2,-2);
    \draw[-latex] (-2,-2) -- (-1,-1);
    \draw (-2,2) -- (-3,2.5);
    \draw (-2,-2) -- (-3,-2.5);
    \draw (-3,1) -- (-3.7,1.2);
    \draw (-3,-1) -- (-3.7,-1.2);
    \begin{scope}[red,decoration={markings,mark=at position 0.15 with {\arrow{>}},mark=at position 0.85 with {\arrow{<}}}]
        \draw[postaction={decorate}] (-3,0) node[above left] {\footnotesize $\ell_2$} -- (-1,0)  node[above right] {\footnotesize $\ell_5$};
        \draw[postaction={decorate}] (-2.5,1.5) node[above] {\footnotesize $\ell_1$} -- (-1.5,-1.5) node[below] {\footnotesize $\ell_4$};
        \draw[postaction={decorate}] (-2.5,-1.5) node[below] {\footnotesize $\ell_3$} -- (-1.5,1.5) node[above] {\footnotesize $\ell_6$};
    \end{scope}
    \begin{scope}[red,decoration={markings,mark=at position 0.5 with {\arrow{<}}}]
        \draw[postaction={decorate}] (-3,0) -- (-3.7,-0.1);
        \draw[postaction={decorate}] (-2.5,1.5) -- (-3.5,1.9);
        \draw[postaction={decorate}] (-2.5,-1.5) -- (-3.5,-1.9);
    \end{scope}
    \filldraw[color=red,fill=white] (-2,0) ellipse (0.4 and 1) node {$\alpha_f$};

    \draw[-latex] (1,-1) --  (1,1);
    \draw[-latex] (1,1) -- (2,2);
    \draw[-latex] (2,2) -- (3,1);
    \draw[-latex] (3,1) -- (3,-1);
    \draw[-latex] (3,-1) -- (2,-2);
    \draw[-latex] (2,-2) -- (1,-1);
    \draw (2,2) -- (3,2.5);
    \draw (2,-2) -- (3,-2.5);
    \draw (3,1) -- (3.7,1.2);
    \draw (3,-1) -- (3.7,-1.2);
    \begin{scope}[red,decoration={markings,mark=at position 0.15 with {\arrow{<}},mark=at position 0.85 with {\arrow{>}}}]
        \draw[postaction={decorate}] (3,0) node[above right] {\footnotesize $\ell_2$} -- (1,0)  node[above left] {\footnotesize $\ell_5$};
        \draw[postaction={decorate}] (2.5,1.5) node[above] {\footnotesize $\ell_1$} -- (1.5,-1.5) node[below] {\footnotesize $\ell_4$};
        \draw[postaction={decorate}] (2.5,-1.5) node[below] {\footnotesize $\ell_3$} -- (1.5,1.5) node[above] {\footnotesize $\ell_6$};
    \end{scope}
    \begin{scope}[red,decoration={markings,mark=at position 0.5 with {\arrow{>}}}]
        \draw[postaction={decorate}] (3,0) -- (3.7,-0.1);
        \draw[postaction={decorate}] (2.5,1.5) -- (3.5,1.9);
        \draw[postaction={decorate}] (2.5,-1.5) -- (3.5,-1.9);
    \end{scope}
    \filldraw[color=red,fill=white] (2,0) ellipse (0.4 and 1) node {$\alpha_f^*$};
\end{tikzpicture}
    \caption{Two polyhedra in a polytope decomposition of a 3-manifold with a common face $f$. The ``colors'' $\ell_i\equiv \ell(e_i)\in \mathrm{Irr}(\mathcal{C})$ are the simple objects assigned to the edges $e_i\in \partial f$ with the orientation indicated by arrows. The graphs dual to the polyhedra are shown in red. Gluing the polyhedra along the common face $f$ corresponds to a summation over the basis elements $\alpha_f\in \mathrm{Hom}_\mathcal{C}(
    \ell(e_1)\otimes \ell(e_2)\otimes \ldots,\mathbf{1})$ assigned to the vertex corresponding to $f$ of the graph on the left. The vertex corresponding to $f$ of the right graph is assigned with the dual basis element $\alpha^*_f\in \mathrm{Hom}_\mathcal{C}(
    \ldots \otimes \ell(e_2)^*\otimes \ell(e_1)^*,\mathbf{1})= \mathrm{Hom}_\mathcal{C}(
    \ell(e_1)\otimes \ell(e_2)\otimes \ldots,\mathbf{1})^*$.} 
    \label{fig:face-gluing}
\end{figure}

Let now $M$ be a closed oriented 3-manifold with a polytope decomposition. The partition function of the Turaev-Viro TQFT on $M$ then reads
\begin{equation}
    Z_\text{TV}(M):=\mathcal{D}^{-2|\mathrm{Vertices}(M)|}\sum_\ell\sum_{\alpha} \prod_{P\in \mathrm{Polyhedra}(M) }Z(P,\ell|_P,\alpha|_P)\prod_{e\in \mathrm{Edges}(M)}d_{\ell(e)}
    \label{TV-poly-closed}
\end{equation}
where $\mathcal{D}^2:=\sum_{\ell\in \mathrm{Irr}(\mathcal{C})}d_\ell^2$. The external sum is performed over $\ell:\mathrm{Edges}(M)\rightarrow \mathrm{Irr}(\mathcal{C})$ -- the assignments of simple objects to oriented edges of the polytope decomposition.  The internal sum is performed over $\alpha=\{\alpha_f\}_f$ -- the assignments of basis morphisms to the oriented faces $f\in \mathrm{Faces}(P)$ of the polytope decomposition. To a face with opposite orientation one assigns a dual basis element $\alpha_{\bar{f}}=\alpha_f^*$. The assignments of the morphisms $\alpha|_P$ for each polyhedron $P$ is then determined by $\alpha$ according to the orientation induced from the orientation of $P$ itself: see Figure \ref{fig:face-gluing}. One can argue that the result is independent of the choice of the polytope decomposition of $M$.

Formula (\ref{TV-poly-closed}) is extended to manifolds with boundaries as follows:
\begin{multline}
    Z_\text{TV}(M):=\mathcal{D}^{-2|\mathrm{Vertices}(M\setminus \partial M)|-|\mathrm{Vertices}(\partial M)|}\times 
    \\
    \sum_\ell\sum_{\alpha} \prod_{P\in \mathrm{Polyhedra}(M) }Z(P,\ell|_P,\alpha|_P)\prod_{e\in \mathrm{Edges}(M\setminus \partial M)}d_{\ell(e)} \prod_{e\in \mathrm{Edges}(\partial M)}\sqrt{d_{\ell(e)}}.
    \label{TV-poly-with-boundary}
\end{multline}
In the summation, $\ell$ runs over the colorings of all the edges, including the ones at the boundary, while $\alpha$ runs only over the colorings of internal faces. The faces belonging to the boundary $\partial M$ remain unassigned, so, rather than being a complex number, the result can be interpreted as an element of a certain vector space associated to the boundary equipped with its polytope decomposition:
\begin{equation}
    Z_\text{TV}(M) \in \bigoplus_{\ell|_{\partial M}}\bigotimes_{f\in\mathrm{Faces}(\partial M)} H(f,\ell|_f)
\end{equation}
where the tensor product is over the boundary faces and
\begin{equation}
    H(f,\ell):=\mathrm{Hom}_\mathcal{C}(\mathbf{1},\ell(e_1)\otimes\ell(e_2)\otimes\ldots)
    \label{face-hilbert-space}
\end{equation}
with $e_i\in \partial f$ oriented according to the orientatation of $f\subset \partial M$ induced from $M$. The result, however, is always inside a certain subspace:
\begin{equation}
    Z_\text{TV}(M)\in H_\text{TV}(\partial M):=\mathrm{Im}(Z_{TV}(\partial M\times I ))\subset  \bigoplus_{\ell|_{\partial M}}\bigotimes_{f\in\mathrm{Faces}(\partial M)} H(f,\ell)
\end{equation}
which can be identified with the image of
\begin{equation}
    Z_{TV}(\partial M\times I):\bigoplus_{\ell|_{\partial M}}\bigotimes_{f\in\mathrm{Faces}(\partial M)} H(f,\ell)
    \longrightarrow \bigoplus_{\ell|_{\partial M}}\bigotimes_{f\in\mathrm{Faces}(\partial M)} H(f,\ell)
    \label{TV-projector}
\end{equation}
where $\partial M\times I$ is given a polytope decomposition by extending the polygons $f\in \partial M$ to polyhedra $f\times I$. The map $Z_{TV}(\partial M\times I)$ satisfies the property 
\begin{equation}
    Z_{TV}(\partial M\times I)\circ Z_{TV}(\partial M\times I)=Z_{TV}(\partial M\times I)
\end{equation}
and thus can be understood as the projector on the subspace $H_\text{TV}(\partial M)$. Moreover, one can show that for two different polygonal decompositions of $\partial M$ the corresponding vector spaces $H_\text{TV}(\partial M)$ are canonically isomorphic. The vector space $H_\text{TV}(\partial M)$ plays the role of the Hilbert space of the Turaev-Viro TQFT on the spatial slice $\partial M$.

\section{UV to IR}\label{sec:uv_to_ir}

In this section we give a proof of Conjecture \eqref{eq:conj} in the context of Levin-Wen models. We begin in Section \ref{sec:LWground} with a brief review of the Levin-Wen model groundstates building on \cite{kirillov2011stringnetmodelturaevviroinvariants,kirillov2010turaevviroinvariantsextendedtqft} and then proceed computing the 4-partite entanglement invariant associated to it. We then proceed in Section \ref{sec:extraction} and extract from its signal the corresponding TQFT partition function.

\subsection{Levin-Wen groundstate and its 4-partite entanglement}\label{sec:LWground}

For the UV theory we consider the string-net model of Levin-Wen \cite{Levin:2004mi}. The ground states are described as wave functions on the space of \textit{string-net} configurations (that is configurations of strings with possible junctions) in a 2-dimensional space, possibly with some boundary conditions. Moreover, the ground state wave functions have equal values on the configurations related by certain local moves. As in the Turaev-Viro TQFT, the input data of the theory is a spherical fusion category. Mathematically, the string-net configurations are simply the graphs with edges labeled by simple objects of the category and vertices by the basis morphisms, as in the standard graphical calculus for fusion categories. The moves then correspond to the identities in the fusion category. The model can have a particular lattice implementation (e.g. honeycomb), in which the strings (graph edges) must be supported on the fixed lattice edges. Moreover, one can engineer a particular lattice Hamiltonian for which the ground states are described by such wave functions. Neither the particular structure of the lattice nor the Hamiltonian will be important for our analysis. 

On the level of the ground states (i.e. in the IR limit) the string-net model of Levin-Wen is described by the Turaev-Viro TQFT, as was argued in particular in \cite{Kadar:2009fs,Koenig:2010uua,kirillov2011stringnetmodelturaevviroinvariants}. Here we will show how this implies a certain relationship between the multi-partite entanglement measures of the ground states in the string-net model and partition functions of the Turaev-Viro TQFT on arbitrary 3-manifolds. This, in a sense, generalizes the relation between the $S^3$ partition function of the TQFT and the von Neumann entanglement entropy obtained in \cite{Kitaev:2005dm,levin2006detecting}.

We will primarily focus on the 4-partite entanglement measures of the ground state on the plane with respect to its splitting into four regions as indicated in Figure \ref{fig:tetrahedron} (a). The generalization to arbitrary splittings is straightforward. As in \cite{levin2006detecting} one can consider string configurations independently in the 4 regions. This gives an interpretation of the ground state on the plane as a state in the tensor product of 4 Hilbert spaces associated with the regions. With such a setup, one can then consider a 4-partite entanglement measure corresponding to an arbitrary 4-colored bi-partite graph, as reviewed in Section \ref{sec:multi-entropy}. The 4 colors of the edges correspond to the 4 regions, as illustrated in Figure \ref{fig:tetrahedron} (c).

\begin{figure}[ht!]
    \centering
    \begin{subfigure}{0.3\textwidth}
        \centering
        \begin{tikzpicture}
            \draw (0,0) circle (1.5);
            \draw (0,0) -- (90:1.5);
            \draw (0,0) -- (-30:1.5);
            \draw (0,0) -- (210:1.5);

            \draw (0.7,0.3) node {$A$};
            \draw (-0.7,0.3) node {$B$};
            \draw (0,-0.7) node {$C$};
            \draw (-1.5,1.5) node {$D$};

\end{tikzpicture}
\caption{}
\end{subfigure}
    \quad
    \begin{subfigure}{0.3\textwidth}
        \centering
        \begin{tikzpicture}
             \draw (0,0) -- (2.3,-0.5);
            \draw (2.3,-0.5) -- (3.2,0.5);
            \draw (3.2,0.5) -- (1.6,2.5);
            \draw (0,0) -- (1.6,2.5);
            \draw (2.3,-0.5) -- (1.6,2.5);
            \draw[dashed] (0,0) -- (3.2,0.5);
        \end{tikzpicture}
        \caption{}
    \end{subfigure}
    \quad
    \begin{subfigure}{0.3\textwidth}
        \centering
        \begin{tikzpicture}
             
             \draw[TealBlue,very thick] (0,0) -- (0:1.3) node[right] {$A$};
             \draw[Maroon,very thick] (0,0) -- (90:1.3) node[above] {$B$};
             \draw[OliveGreen,very thick] (0,0) -- (180:1.3) node[left] {$C$};
             \draw[RoyalPurple,very thick] (0,0) -- (270:1.3) node[below] {$D$};

             \draw[fill=black] (0,0) circle (0.15);
        \end{tikzpicture}
        \caption{}
    \end{subfigure}
    \caption{}
    \label{fig:tetrahedron}
\end{figure}

As in \cite{levin2006detecting}, we do not allow strings to go to infinity in the plane, thus the plane can be effectively considered as a 2-sphere. Topologically, it can be understood as the boundary of the tetrahedron shown in Figure \ref{fig:tetrahedron} (b), with the 4 regions being its 4 triangular faces. Let us denote this tetrahedron by $T_\text{IR}$. Consider a single triangle corresponding to one of the four regions on the plane. The edges of the Levin-Wen lattice model intersect its boundary at certain places. Let us subdivide the edges of the triangle into smaller intervals such that each interval intersects the lattice at exactly one point. The space of the ground states in the string-net model on the triangle with the boundary condition that fixes labels of all the strings passing through the boundary then can be identified with (\ref{face-hilbert-space}) \cite{kirillov2011stringnetmodelturaevviroinvariants}, for $f$ being the polygon obtained from the triangle by the subdivision of its edges as above. See Figure \ref{fig:subdivided-triangle}. Let us denote the \textit{polyhedron} obtained by subdividing all the edges of the tetrahedron by $T_\text{UV}$. There $\alpha_f$ stands for a basis ``input'' string-net configuration. We will refer to the number of intervals in the subdivision of the edge $e\in T_\text{IR}$ as its \textit{length} and denote it by $L(e)\in \mathbb{Z}_{\geq 1}$.

\begin{figure}[ht!]
    \centering
        \centering
        \begin{tikzpicture}[scale=1.2]
            \draw (-3,0) -- (0,6) -- (6,0) -- (-3,0);
            \begin{scope}[decoration={markings,mark=at position 0.2 with {\arrow{>}}}]
            \foreach \x in {1,2,...,7} {
                \draw[fill=black] ({-3*\x/7},{6*(1-\x/7)}) circle (0.03);
                \draw[red,postaction={decorate}] ({-3*(\x-1/2)/7},{6*(1-(\x-1/2)/7)}) node[left] {$\ell_\x'$} -- (1,2);
            }
            \foreach \x in {1,2,...,10} {
                \draw[fill=black] ({(-3)*(1-\x/10)+6*\x/10},0) circle (0.03);
                \draw[red,postaction={decorate}] ({(-3)*(1-(\x-1/2)/10)+6*(\x-1/2)/10},0) node[below] {$\ell_{\x}''$} -- (1,2);
            }
            \foreach \x in {1,2,...,9} {
                \draw[fill=black] ({6*(1-\x/9)},{6*\x/9}) circle (0.03);
                \draw[red,postaction={decorate}] ({6*(1-(\x-1/2)/9)},{6*(\x-1/2)/9}) node[right] {$\ell_{\x}'''$} -- (1,2);
            }
            \end{scope}
            \filldraw[color=red,fill=white] (1,2) circle (1.2) node {$\alpha_f$};
        \end{tikzpicture}
        \caption{}
        \label{fig:subdivided-triangle}
\end{figure}

Since the ground state on 2-sphere is unique (up to scalar factor), it can be realized as the state given by the TQFT partition function on a ball, such as the solid version of the  tetrahedron considered above. It is thus provided by the formula (\ref{TV-poly-with-boundary}) in the case when $M$ consists of a single polyhedron $T_\text{UV}$:
\begin{multline}
    Z_\text{TV}(T_\text{UV})=\mathcal{D}^{-|\mathrm{Vertices}(T_\text{UV})|}
    \sum_\ell Z(T_\text{UV},\ell,\cdot) \prod_{e\in \mathrm{Edges}(T_\text{UV})}\sqrt{d_{\ell(e)}}\\
    \in \bigoplus_{\ell}\bigotimes_{f\in\mathrm{Faces}(T_\text{UV})} H(f,\ell|_f)
    \subset \bigotimes_{f\in\mathrm{Faces}(T_\text{UV})}\bigoplus_{\ell_f} H(f,\ell_f).
    \label{Z-TUV}
\end{multline}
The last inclusion is into a large space, where the direct sums over the edge colors are performed independently for each face. This allows us to consider 
\begin{equation}
    |\psi\rangle= {\mathcal{D}^{|\mathrm{Vertices}(T_\text{UV})|}\cdot} Z_\text{TV}(T_\text{UV})
\end{equation}
as a state in the tensor product of four Hilbert spaces corresponding to the four faces. {  Note that the overall normalization factor $\mathcal{D}^{|\mathrm{Vertices}(T_\text{UV})|}$ is introduced for convenience only, as it drops out from the signal, which is independent of the normalization of the state.} Its Hermitian conjugate is provided by $\overline{{T}_{UV}}$, the same tetrahedron but with opposite orientation:
\begin{equation}
    \langle\psi|= {  \mathcal{D}^{|\mathrm{Vertices}(T_\text{UV})|}\cdot} Z_\text{TV}(\overline{T_\text{UV}}).
\end{equation}
In this setup we can use (\ref{Z-TUV}) as an input for a 4-partite entanglement multi-invariant  $\mathcal{Z}(\Gamma;\cdot)$ corresponding to an arbitrary 4-colored bipartite graph $\Gamma$, reviewed in Section \ref{sec:multi-entropy}. The \textit{unnormalized} multi-invarian is then very similar to (\ref{TV-poly-closed}) with $M=M_\text{UV}$ being the oriented \textit{polytope complex} that consists of copies of $T_{UV}$ and $\overline{{T}_{UV}}$ ($T_{UV}$ with reversed orientation) glued with identity map along their faces according to the graph $\Gamma$:
\begin{equation}
    \mathcal{Z}(\Gamma;|\psi\rangle)=\sum_\ell\sum_{\alpha} \prod_{P\in \mathrm{Polyhedra}(M_\text{UV}) }Z(P,\ell|_P,\alpha|_P)\prod_{e\in \mathrm{Edges}(M_\text{UV})}d_{\ell(e)}^{n({e})}.
\end{equation}
Apart from the overall factor, the only difference from (\ref{TV-poly-closed}) is in the exponents of powers of quantum dimensions that appear as weights of in sum over $\ell$: here $n(e)\in \mathbb{Z}_{\geq 1}$ is half the number of polyhedra that have $e$ as a common edge, which is in general different from 1. In the case when $d_\ell=1$ for all simple objects (i.e. the input fusion category is of the form $\mathrm{Vec}_G^\omega$ and thus the TQFT is of Dijkgraaf-Witten kind), this mismatch disappears. Below we argue that in general it can be reduced to an overall factor in the limit of large lengthes $L(e)\gg 1$.

\begin{figure}[ht!]
        \centering
        \begin{tikzpicture}[scale=1.2]
            \draw (-3,0) -- (0,6) -- (6,0) -- (-3,0);
            \begin{scope}[decoration={markings,mark=at position 0.2 with {\arrow{>}}}]
            \foreach \x in {1,2,...,7} {
                \draw[fill=black] ({-3*\x/7},{6*(1-\x/7)}) circle (0.03);
                \draw[red,postaction={decorate}] ({-3*(\x-1/2)/7},{6*(1-(\x-1/2)/7)}) node[left] {$\ell_\x'$} -- (-0.5,2.2);
            }
            \foreach \x in {1,2,...,10} {
                \draw[fill=black] ({(-3)*(1-\x/10)+6*\x/10},0) circle (0.03);
                \draw[red,postaction={decorate}] ({(-3)*(1-(\x-1/2)/10)+6*(\x-1/2)/10},0) node[below] {$\ell_{\x}''$} -- (1,1);
            }
            \foreach \x in {1,2,...,9} {
                \draw[fill=black] ({6*(1-\x/9)},{6*\x/9}) circle (0.03);
                \draw[red,postaction={decorate}] ({6*(1-(\x-1/2)/9)},{6*(\x-1/2)/9}) node[right] {$\ell_{\x}'''$} -- (2,2.5);
            }
            \end{scope}
            \begin{scope}[red,decoration={markings,mark=at position 0.5 with {\arrow{>}}}]
                \draw[postaction={decorate}] (-0.5,2.2) -- (0.8,2.2) node[midway,above] {$\hat\ell'$};
            \end{scope}
            \begin{scope}[red,decoration={markings,mark=at position 0.5 with {\arrow{>}}}]
                \draw[postaction={decorate}] (1,1) -- (0.8,2.2) node[midway,right] {$\hat\ell''$};
            \end{scope}
            \begin{scope}[red,decoration={markings,mark=at position 0.5 with {\arrow{>}}}]
                \draw[postaction={decorate}] (2,2.5) -- (0.8,2.2) node[midway,above] {$\hat\ell'''$};
            \end{scope}
            \filldraw[color=red,fill=white] (-0.5,2.2) circle (0.4) node {$\alpha_f'$};
            \filldraw[color=red,fill=white] (1,1) circle (0.4) node {$\alpha_f''$};
            \filldraw[color=red,fill=white] (2,2.5) circle (0.4) node {$\alpha_f'''$};
            \filldraw[color=red,fill=white] (0.8,2.2) circle (0.4) node {$\hat{\alpha}_f$};
        \end{tikzpicture}
    \caption{}
    \label{fig:triangle-uv-to-ir}
\end{figure}

First, let us decompose the basis morphism $\alpha_f$ corresponding to the faces (see Figure \ref{fig:subdivided-triangle}) into the septuples $(\alpha'_f,\alpha_f'',\alpha_f''',\hat\alpha;\hat{\ell}',\hat{\ell}'',\hat{\ell}''')$ where the first four elements are basis morphisms and the last three elements are simple objects in the graph/string-net shown in Figure \ref{fig:triangle-uv-to-ir}. Let $M_\text{IR}$ be the polytope complex obtained by gluing the copies of $T_{IR}$, instead of $T_{UV}$, according to the same graph $\Gamma$. That is, $M_\text{IR}$ is obtained from $M_\text{UV}$ by removing all the extra vertices that were introduced by subdivision of the edges of the tetrahedron $T_\text{IR}$. Using the identity illustrated in Figure \ref{fig:common-edge} (cf. \cite{Fuchs:2002cm}) one can perform partial summations over the colorings $\ell$: 
\begin{multline}
    \mathcal{Z}(\Gamma;|\psi\rangle)=\\
    \sum_{\hat\ell}\sum_{\hat\alpha} \prod_{P\in \mathrm{Polyhedra}(M_\text{IR}) }Z(P,\hat\ell|_P,\hat\alpha|_P)\prod_{e\in \mathrm{Edges}(M_\text{IR})}\left(
    \sum_{\ell\in \mathrm{Irr}(\mathcal{C})^{L(e)}}N_{\ell_1\ell_2\ldots \ell_{L(e)}}^{\hat\ell(e)}
    \prod_{i=1}^{L(e)}d_{\ell_i}^{n(e)}
    \right),
\end{multline}
where $N_{\ell_1 \ldots \ell_{L}}^{\hat\ell}:=\dim\,\mathrm{Hom}_\mathcal{C}(\ell_1\otimes \ldots \otimes \ell_{L},\hat\ell)$ are fusion coefficients.

\begin{figure}[ht!]
    \centering
    \begin{tikzpicture}
        \draw (0,0) -- (0,6);
        \draw (0,0) -- (-1,0.3);
        \draw (0,0) -- (1,0.3);
        \draw (0,6) -- (1,7);
        \draw (0,6) -- (-1,7);
        \begin{scope}[decoration={markings,mark=at position 0.5 with {\arrow{>}}}]
            \foreach \x in {1,2,...,8} {
                \draw[red,postaction={decorate}] (0,{6*(1-(\x-1/2)/8)}) 
                -- (2,4);
                \draw[red,postaction={decorate}] (-2,4) -- (0,{6*(1-(\x-1/2)/8)});
                \draw[fill=black] (0,6*\x/8) circle (0.03);
            }
            \draw[fill=black] (0,0) circle (0.03);
            \draw[red,postaction={decorate}] (-3.5,4.7) -- (-2,4) node[midway,above] {$\hat{\ell}_1$};
            \draw[red,postaction={decorate}] (2,4) -- (3.5,4.7) node[midway,above] {$\hat{\ell}_2$};
        \end{scope}
        \filldraw[color=red,fill=white] (-2,4) ellipse (0.4 and 1) node {$\alpha_{f_1}'$};
        \filldraw[color=red,fill=white] (2,4) ellipse (0.4 and 1) node {$\alpha_{f_2}'$};
    \end{tikzpicture}
    \begin{tikzpicture}
        \draw (0,0) -- (0,6);
        \draw (0,0) -- (-1,0.3);
        \draw (0,0) -- (1,0.3);
        \draw (0,6) -- (1,7);
        \draw (0,6) -- (-1,7);
        \draw[fill=black] (0,0) circle (0.03);
        \draw[fill=black] (0,6) circle (0.03);
        \begin{scope}[decoration={markings,mark=at position 0.5 with {\arrow{>}}}]
            \draw[red,postaction={decorate}] (-1.5,4) -- (0,3) node[midway,above] {$\hat{\ell}_1$};
            \draw[red,postaction={decorate}] (0,3) -- (1.5,4) node[midway,above] {$\hat{\ell}_1$};
        \end{scope}
        \draw (-3,3) node {$=\delta_{\hat{\ell}_1,\hat{\ell}_2}\,\delta_{\alpha_{f_2}',\alpha_{f_1}'^*}$};
    \end{tikzpicture}
    \caption{}
    \label{fig:common-edge}
\end{figure}

In the next step we will use the fact that for a unitary spherical fusion category
    \begin{equation}
     \sum_{\ell\in \mathrm{Irr}(\mathcal{C})^{L}}N_{\ell_1\ell_2\ldots \ell_{L}}^{\hat\ell}
    \prod_{i=1}^{L}d_{\ell_i}^{n}=\frac{\left(\sum_{\ell\in\mathrm{Irr}(\mathcal{C})}d_\ell^{n+1}\right)^L\,d_{\hat\ell}}{\mathcal{D}^2}\left(1+O(e^{-c_n\,L})\right),\quad L\gg 1,
    \label{fusion-sum-limit}
\end{equation}
for some $c_n>0$. The statement follows from the Frobenius-Perron theorem applied to the matrices of fusion coefficients, as we argue in Appendix \ref{app:fusion}. There we also  provide an alternative more explicit proof for the case when $\mathcal{C}$ is a modular tensor category (cf. \cite{Flammia:2009axf}), using the Verlinde formula.

 When $n=1$ the exponentially suppressed correction vanishes identically, and the formula becomes an exact, well-known one:
\begin{equation}
     \sum_{\ell\in \mathrm{Irr}(\mathcal{C})^{L}}N_{\ell_1\ell_2\ldots \ell_{L}}^{\hat\ell}
    \prod_{i=1}^{L}d_{\ell_i}=\mathcal{D}^{2L-2}{d_{\hat\ell}}.
\end{equation}
From this it also automatically follows that (\ref{fusion-sum-limit}) holds with vanishing correction term if $d_\ell=1,\;\forall\ell$, that is when $\mathcal{C}\cong \mathrm{Vec}_G^\omega$ and the Turaev-Viro TQFT is of Dijkgraaf-Witten type. 

Let $M_\text{IR}$ (which is, of course, homeomorphic to $M_\text{UV}$) be a 3-manifold. For this to be the case, it is necessary and sufficient that the link of any vertex in  $M_\text{IR}$ is homeomorphic to a sphere. On the level of the graph $\Gamma$, this is the condition that all 3-colored connected subgraphs are planar. Using (\ref{fusion-sum-limit}), we thus have
\begin{multline}
    \mathcal{Z}(\Gamma;|\psi\rangle)=\left(1+O(e^{-c\,L_\mathrm{min}})\right)\,Z_\text{TV}(M_\text{IR})\times\\
   \mathcal{D}^{2|\mathrm{Verices}(M_\text{IR})|-2|\mathrm{Edges}(M_\text{IR})|}\prod_{e\in \mathrm{Edges}(M_\text{IR})}\mathcal{D}_{(n(e))}^{L(e)}
    \label{TV-ent}
\end{multline}
where $L_\mathrm{min}:=\min_{e\in \mathrm{Edges}(T_\text{IR})}L(e)$, $c=\min_{e\in\mathrm{Edges}(T_\text{IR})}c_{n(e)}$, and
\begin{equation}
 \mathcal{D}_{(n)}:=\sum_{\ell\in\mathrm{Irr}(\mathcal{C})}d_\ell^{n+1}.
\end{equation}
Note that $\mathcal{D}_{(1)}\equiv \mathcal{D}^2$.

One can rewrite the factors in (\ref{TV-ent}) in terms of the data of the graph $\Gamma$:
\begin{multline}
    \mathcal{Z}(\Gamma;|\psi\rangle)=\left(1+O(e^{-c\,L_\mathrm{min}})\right)\,Z_\text{TV}(M_\text{IR})\times\\
   \mathcal{D}^{2N_{3}(\Gamma)}\prod_{(e,n)\in \mathrm{Edges}(T_\text{IR})\times \mathbb{Z}_{\geq 1}}\left(\mathcal{D}_{(n)}^{L(e)}/\mathcal{D}^2\right)^{N_{2:e,n}(\Gamma)}
    \label{TV-ent-graph}
\end{multline}
where $N_3(\Gamma)$ is the number of 3-colored subgraphs of $\Gamma$ and $N_{2:e,n}(\Gamma)$ is the number of 2-colored loops of length $2n$ with the 2 colors corresponding to the edge $e\in \mathrm{Edges}(T_\text{IR})$. Using this expression, we show below that the absolute value of the TQFT partition function $Z(M)\equiv Z_\text{TV}(M_\text{IR})=Z_\text{TV}(M_\text{UV})$ can be extracted as one of the signals (reviewed in Section \ref{construction_signal}) that captures genuine 4-partite entanglement. In Appendix \ref{app:alt} we provide an alternative method to extract $Z(M)$ itself in terms of $\mathcal{Z}(\Gamma;|\psi\rangle)$ and $\mathcal{Z}(\Gamma^\lambda;|\psi\rangle)$ for some universal ($\Gamma$-independent) collection of graphs.

We remark that one can in principle consider in a similar way an arbitrary subdivison of a plane into connected regions, which would correspond to considering an arbitrary (topological) polyhedron instead of the tetrahedron. The arguments above extend to such a more general setting. In particular, for a subdivision into two connected regions, one would consider a \textit{dihedron}. The standard Rényi entropies then correspond to the gluing of the copies of this dihedron and its orientation reversal along the common faces. Such a gluing always produces $S^3$. From this point of view, it is clear that Rényi entropies do not provide any information about the IR TQFT in addition to the ordinary topological entropy of \cite{Kitaev:2005dm,levin2006detecting} (which is given by $2\log \mathcal{D}=-\log Z(S^3)$), as was indeed observed in \cite{Flammia:2009axf}.  Similarly, a subdivision into 3 connected regions leads to at most a connected sum of copies of $S^2\times S^1$. The partition function in this case can also be expressed through the one on $S^3$.

\subsection{Extracting TQFT contribution}\label{sec:extraction}
In the previous section we computed a general $4$-partite multi-invariant ${\cal Z}(\sigma_1,\sigma_2,\sigma_3,\sigma_4;\cdot)$ that is a GEM for some manifold $M$. Let the regions associated to the permutation arguments are $A, B, C$ and $D$ respectively. The multi-invariant yields the partition function of the TQFT on $M$ but it is  multiplied by certain other undesirable UV contributions as shown in equation \eqref{TV-ent-graph}.  
These multiplicative contributions are
\begin{align}\label{uv_contribution}
    \mathcal{D}^{2N_{3}(\Gamma)}\prod_{(e,n)}\left(\mathcal{D}_{(n)}^{L(e)}/\mathcal{D}^2\right)^{N_{2:e,n}(\Gamma)}={\cal D}_{(1)}^{-N_0(\Gamma)} \prod_{(i,j)} \prod_{\rm loops}{\cal D}_{(n_{\rm loop})}^{\ell_{ij}}.
\end{align}
On the right-hand side, we have translated the multiplicative factor in the language of the multi-invariant graph. Using the Euler relation  \eqref{relation-euler} for $3$-manifolds we have combined the contribution of ${\cal D}^2={\cal D}_{(1)}$ from both factors. 
We have also organized the product of ${\cal D}_{(n)}$ over the edges of triangulation according to the type of loop in the multi-invariant graph. Here  $(i,j)$ denotes a distinct pair of colors. 
For a given $(i,j)$, the 2-colored subgraph of a multi-invariant with those colors consists of disjoint loops. The second product is over those loops with $n_{\rm loop}$ being the length of a given loop. In order to extract the topological partition function, we would like to develop a scheme to remove these UV contributions. In fact, we will be able to formally identity this contribution as coming from certain (fictitious)  bi-partite states localized at the boundaries of the regions. This interpretation makes it clear that this contribution vanishes for any $4$-partite signal. 

We would like to consider other multi-invariants of the same family as ${\cal Z}$ and use them to remove the UV contributions. An example of such  multi-invariant is ${\cal Z}(\sigma_1,\sigma_1,\sigma_3, \sigma_4;\cdot)$. Because the permutation element for party $A$ and $B$ are the same, this multi-invariant is in fact tri-partite. It treats parties $A$ and $B$ as the same party. It is a tri-partite multi-invariant where parties $A$ and $B$ are treated as a single party because the permutation elements corresponding to them are the same. Each invariant of the same family is specified by specifying a map from the set of parties $\{A, B, C,D\}$ to the 
permutation set $\Sigma=\{\sigma_1, \ldots, \sigma_4\}$. The multi-invariant graph for each member of the family is understood as follows. For ${\cal Z}(\sigma_1, \sigma_2, \sigma_3, \sigma_4;\cdot)$ has $4$  \emph{types} of edges labeled by permutation elements. This graph with $4$ types of edges is $\Gamma$. For multi-invariant ${\cal Z}(\sigma_1, \sigma_2, \sigma_3, \sigma_4;\cdot)$, we say that the edges of the types $\sigma_1, \sigma_2, \sigma_3$ and $\sigma_4$ are colored with \emph{colors} $A, B, C$ and $D$ respectively. 
The graph of multi-invariant is ${\cal Z}(\sigma_1,\sigma_1,\sigma_3, \sigma_4;\cdot)$ is  also  $\Gamma$ but now the edges of the type $\sigma_1$ is colored with two colors $A$ and $B$, edges of the type $\sigma_3$ and $\sigma_4$ are colored with colors $C$ and $D$ respectively.  While the edges of the type $\sigma_2$ are uncolored or absent. Note that a given color can only be used for one type of edge. 

Now we will rewrite the UV contribution in a form that is valid for all the multi-invariants of the same family $\Sigma$. In particular, in a way that makes sense even for multi-invariants that are less than $4$-partite. We will do so by looking at the contribution from certain subgraphs of $\Gamma$. This contribution depends on the coloring of the edges of $\Gamma$.

Consider the contribution due to the subgraph with edges of two types, say $\alpha, \beta$. If both these types have a single color, then its contribution is
\begin{align}\label{boundary_length}
    \prod_{\rm loops} {\cal D}_{n_{\rm loop}}^{\ell_{r(\alpha), r(\beta)}}.
\end{align}
Here the product is over the number of connected components of the subgraph. Each connected component is a loop that alternates between $\alpha$ and $\beta$ and $n_{\rm loop}$ is the number of $\alpha$ (equivalently $\beta$) edges of the loop.  The notation $\ell_{a,b}$ stands for the length of the boundary separating regions $a$ and $b$.  If the edges $\alpha$ are colored with two colors, say $A$ and $D$, and $\beta$ is colored with, say $B$ and $C$, then there is contribution coming from multiple pairs of colors $(A, B), (A, C), (D, B)$ and $(D,C)$. Each of these contributions comes with a different exponent of ${\cal D}_{n_{\rm loop}}$ e.g. the contribution due to the pair $(D,B)$ comes with the exponent $\ell_{D,B}$ and so on. Interestingly the formula \eqref{boundary_length} takes into account all such contributions.

There is also a contribution associated to a subgraph of consisting of edges of a single type, say  $\alpha$. It consists of disconnected components each consisting of a single $\alpha$ type edge. 
If edges of type $\alpha$ are colored with a single color then this contribution is just $1$. 
We do get a non-trivial contribution when $\alpha$ edges are colored with more than one color. If it has two colors, say $A$ and $B$ then the contribution comes because of this pair of colors. It is equal to 
\begin{align}
    \prod_{{\rm edges}} {\cal D}_{(1)}^{\ell_{r(\alpha)}}.
\end{align}
Here $\ell_{a}$ is defined as the length of the marked edge \emph{within} the region $a$. This formula holds even if $\alpha$ is colored by more than $3$ colors.

Finally, there is a contribution ${\cal D}_{(1)}^{-N_0(\Gamma)}$ where $N_0(\Gamma)$ is the total number of vertices of $\Gamma$.

Combining all the contributions, we get a formula that works for general multi-invariant of family $\Sigma$, in particular, even when that multi-invariant is lower-partite. The  contribution in equation \eqref{uv_contribution} is written as
\begin{align}\label{uv-refined}
     \Big(\prod_{\alpha, \beta}\prod_{\rm{loops}} {\cal D}_{(n_{\rm loop})}^{{\ell_{r(\alpha), r(\beta)}}}  \prod_\alpha \prod_{\rm edges} {\cal D}_{(1)}^{\ell_{r(\alpha)}}\Big)\Big( \prod_{\rm vertices}{\cal D}_{(1)}^{-1}\Big).
\end{align}
The first factor is consistent with the contribution of bi-partite states localized along the edges that are entangled across the edge and have the property ${\rm Tr}\rho^n={\cal D}_{(n)}$. The second factor is consistent with the contribution of a single-partite (un-entangled) state with norm $\mathcal D_{(1)}^{-1}$. Of course, the bi-partite state with the above properties may not exist, but this does not matter for our purposes. We will be using lower partite multi-invariants to cancel the 
UV contributions. All the multi-invariants would contribute formally according to the existence of such ``fictitious'' states and can be used to cancel these contributions.

The lower-partite multi-invariant of family $\Sigma$ is obtained by choosing permutations that are not all distinct. The graph of such multi-invariant is a particular subgraph ${\cal Z}(\sigma_1, \sigma_2, \sigma_3, \sigma_4;\cdot)$ generated by edges of $3$ or less number of types. Because the multi-invariant  ${\cal Z}(\sigma_1, \sigma_2, \sigma_3, \sigma_4;\cdot)$ is a GEM, the lower-partite multi-invariant of the same family must also be a GEM but for  a disjoint union of copies of $S^3$. Hence the topological contribution coming from all such lower-partite multi-invariant is $Z(S^3)$. 
We can use the simple signal constructed in Section \ref{construction_signal} to extract the genuine $4$-partite contribution: 
\begin{align}
    {\cal E}(\sigma_1-\sigma_2, \sigma_2-\sigma_1,\sigma_3-\sigma_1,\sigma_4-\sigma_1;\cdot).
    \label{genuine-4-specific}
\end{align}
It has $16$ terms and is equal to 
\begin{equation}
 \log Z(M)+\log \overline{Z({M})}-n_{\mathtt S,\Delta}\log Z(S^3) . 
 \label{signal-specific-result}
\end{equation}
The first to terms are the TQFT contributions from $\mathcal{E}(\sigma_1,\sigma_2,\sigma_3,\sigma_4;\cdot)$ and $\mathcal{E}(\sigma_2,\sigma_1,\sigma_3,\sigma_4;\cdot)$ and we used the fact that the partition function on $\overline{M}$, the orientation reversal of $M$, is $Z(\overline{M})=\overline{Z(M)}$ by unitarity (reflection-positivity). The coefficient $n_{\mathtt S,\Delta}$ in the third term is an integer which gets contribution from connected components of the multi-invariant graphs that appear in the linear combination in \eqref{genuine-4-specific}. In Appendix \ref{app:example} we demonstrate explicitly how the cancellation works in (\ref{genuine-4-specific}) in the example of a graph realizing a Lens space 3-manifold. In that case, $n_{\mathtt S,\Delta}=2$.

The signal is constructed so that it removes the contribution of any state that is less than $4$-partite entanglement. 
Interestingly, we see that the Levin-Wen model does not seem to have any tri-partite entanglement localized at the points that are common to three regions. As a result, we could have used a combination that is not necessarily the signal of genuine $4$-partite entanglement but only a combination that removes bi-partite entanglement. However, other microscopic models that flow to TQFTs at long distances may have such a contribution and so it is better to use the signal of genuine $4$-partite entanglement to remove such a contribution and end up with a purely non-local or long-range entanglement. 

Assuming the cancellation of non-universal UV contributions, any other signal (\ref{signal-general}) in any dimension will produce the result of the same form as in (\ref{signal-specific-result}), with $\log Z(M)$ and $\log \overline{Z(M)}$ contributing with the same coefficients.
Namely, up to exponentially suppressed corrections,
\begin{multline}
    \mathcal{E}(\sum_{i_1}a_{i_1}^{(1)}\sigma_{i_1},\ldots,\sum_{i_{d+1}}a_{i_{d+1}}^{(d+1)}\sigma_{i_{d+1}};|\psi\rangle_H) =\\
        =k_{\mathtt{S}}(\log Z(M)+\log \overline{Z({M})})-n_{\mathtt{S},\Delta}\log Z(S^d).
\end{multline}
This can be seen as follows. The difference between the coefficients in front of $\log Z(M)$ and $\log \overline{Z({M})}$ is given by
\begin{equation}
    \sum_{\xi\in S_{d+1}}(-1)^{\epsilon(\xi)}\,a_{\xi(1)}^{(1)}a_{\xi(2)}^{(2)}\dots  a_{\xi(d+1)}^{(d+1)}=\det (a^{(j)}_i)_{j,i}=0.
\end{equation}
where $\epsilon(\xi)$ is the parity of the permutation $\xi$ of the $q=d+1$ parties. The vanishing of the determinant is implied by the conditions $\sum_ia^{(j)}_i=0,\;\forall j$ in (\ref{signal-general}). Moreover,
\begin{equation}
    k_{\mathtt{S}}= \sum_{\xi\in S_{d+1}}a_{\xi(1)}^{(1)}a_{\xi(2)}^{(2)}\dots  a_{\xi(d+1)}^{(d+1)}.
\end{equation}
However, up to normalization, one can assume $k_{\mathtt{S}}=2$, which would give the conjectural formula (\ref{eq:conj}).

Finally, let us comment on the situation when the graph $\Gamma$ does not represent a 3-manifold. This happens if and only if it contains non-planar 3-colored subgraphs. Suppose this is the case, and some connected components of 3-colored subgraphs have positive genus. In terms of the dual simplicial complex (in which the tetrahedra correspond to the vertices of $\Gamma$), this means that links of some vertices are (triangulated) surfaces $\Sigma_{g_i}$ of positive genus $g_i>0$. This simplicial complex can be understood as a 3-manifold with conical singular points. Namely, such a point has a vicinity homeomorphic to a cone over $\Sigma_{g_i}$, that is $\Sigma_{g_i}\times [0,1)/\Sigma_i\times \{0\}$. Let now $M$ be a non-singular manifold obtained by removing such vicinities. The cost of this operation is introduction of boundaries: $\partial M\cong \sqcup_i \Sigma_{g_i}$. 

We claim that in this case the signal, up to exponentially supressed corrections, gives a linear combination of partition functions $Z_\text{D}$ with ``Dirichlet'' boundary condition imposed on the boundaries :
\begin{equation}
    \log Z_\text{D}(M)+\log\overline{Z_\text{D}(M)}-\sum_{g}n^{(g)}_{\mathtt S,\Delta}\log Z_\text{D}(\Sigma_g\times [0,1])).
\end{equation}
The genera $g$ which appear in the sum with non-zero coefficients are the genera of the surfaces $\Sigma_{g_i}$ at the boundary of $M$. The ``Dirichlet'' boundary condition is the canonical\footnote{It is canonical for a fixed input $\mathcal{C}$. If one considers a different input category  $\mathcal{C}'$ that results in an equivalent TQFT, such a boundary condition in general will be different.} topological boundary condition  for Turaev-Viro TQFT corresponding to the input fusion category $\mathcal{C}$. In terms of the general description of the Turaev-Viro partition function on a manifold with boundary (\ref{TV-poly-with-boundary}) via its polyhedral decomposition, the corresponding boundary state is specified by fixing the boundary coloring to be trivial, that is taking all objects associated to the boundary edges to be $\mathbf{1}$ and all morphisms asssociated to boundary edges to be identities $\mathrm{id}_{\mathbf{1}}\in \mathrm{Hom}_\mathcal{C}(\mathbf{1},\mathbf{1})$. The claim above then follows from the fact that the polyhedral decomposition of $M$  can be on obtained by starting from the original simplicial complex associated to the graph $\Gamma$ and cutting out sufficiently small corners of the tetrahedra containing the conical singularities, see Figure \ref{fig:corner-cut}. In particular, the new triangular faces created by this process constitute a triangulation of the boundary components of $M$. A general coloring of the original simplicial complex (with its edges possibly subdivided) then provides a general coloring of the polyhedral decomposition of $M$ which is trivial at the boundary, and thus corresponds to the canonical topological boundary condition.

\begin{figure}[ht!]
    \centering

       \begin{tikzpicture}[scale=0.7]
             \draw (0,0) -- (5,-1);
              \draw (0,0) -- (2.5,5);
            \draw[dashed] (0,0) -- (5,1);
    \begin{scope}[red,decoration={markings,mark=at position 0.35 with {\arrow{>}}}]
    \draw[postaction={decorate}] (2,-0.4) -- (5,1.5) ;
    \draw[postaction={decorate}] (4,-0.8) -- (5,0.5) ;

    \draw[postaction={decorate}] (1,2) -- (4.5,3) ;

    \draw[postaction={decorate}] (2,4) -- (3.5,4) ;

    \end{scope}

\begin{scope}[red,dashed,decoration={markings,mark=at position 0.5 with {\arrow{<}}}]
    \draw[postaction={decorate}] (2,-0.4) -- (5,0) ;
    \draw[postaction={decorate}] (4,-0.8) -- (5,-0.5) ;

    \draw[postaction={decorate}] (1,2) -- (4.5,4) ;

    \draw[postaction={decorate}] (2,4) -- (3.5,4.5) ;

    \draw[postaction={decorate}] (3,0.6) -- (4.5,2) ;

    \draw[postaction={decorate}] (5,0.7) -- (3,0.6);
   
    \end{scope}

     \draw[fill=black] (0,0) circle (0.05);

    \draw[fill=black] (3,-0.6) circle (0.05);

     \draw[fill=black] (1.5,3) circle (0.05);
    
        \end{tikzpicture}\qquad
         \begin{tikzpicture}[scale=0.7]
             \draw (1,-0.2) -- (5,-1);
              \draw (0.5,1) -- (2.5,5);
            \draw[dashed] (0.3,0.06) -- (5,1);
            \draw (0.5,1) -- (1,-0.2) -- (0.3,0.06) -- (0.5,1);
    \begin{scope}[red,decoration={markings,mark=at position 0.35 with {\arrow{>}}}]
    \draw[postaction={decorate}] (2,-0.4) -- (5,1.5) ;
    \draw[postaction={decorate}] (4,-0.8) -- (5,0.5) ;

    \draw[postaction={decorate}] (1,2) -- (4.5,3) ;

    \draw[postaction={decorate}] (2,4) -- (3.5,4) ;

    \end{scope}

\begin{scope}[red,dashed,decoration={markings,mark=at position 0.5 with {\arrow{<}}}]
    \draw[postaction={decorate}] (2,-0.4) -- (5,0) ;
    \draw[postaction={decorate}] (4,-0.8) -- (5,-0.5) ;

    \draw[postaction={decorate}] (1,2) -- (4.5,4) ;

    \draw[postaction={decorate}] (2,4) -- (3.5,4.5) ;

    \draw[postaction={decorate}] (3,0.6) -- (4.5,2) ;

    \draw[postaction={decorate}] (5,0.7) -- (3,0.6);
   
    \end{scope}

     \draw[fill=black] (0.5,1) circle (0.05);

      \draw[fill=black] (1,-0.2) circle (0.05);

       \draw[fill=black] (0.3,0.06) circle (0.05);

    \draw[fill=black] (3,-0.6) circle (0.05);

     \draw[fill=black] (1.5,3) circle (0.05);

 \draw (-2,2) node {$\rightsquigarrow$};
    
        \end{tikzpicture} 
  
    \caption{Cutting out the corners attached to conical singularities.}
    \label{fig:corner-cut}
\end{figure}

\section{Future directions}\label{sec:generalizations}
In this paper, we have solidified the relationship between ground-state multipartite entanglement and the low-energy TQFT through Conjecture~\eqref{eq:conj}. Our conjecture is broad and applies to a wide range of scenarios. In this work, we verified it in the specific setting of the Levin-Wen string-net model. Natural future directions include exploring the conjecture in other settings. We expect that higher-dimensional topological theories - such as Dijkgraaf-Witten \cite{Dijkgraaf:1989pz} theory and (generalizations of) Crane-Yetter theory \cite{Crane:1993if,Crane:1994ji,douglas2018fusion} - that admit a Turaev-Viro-type state-sum model for the partition function should allow a straightforward verification of Conjecture~\eqref{eq:conj} using the tools developed here.

It would be very interesting to explore Conjecture~\eqref{eq:conj} for chiral topological theories such as Chern-Simons theories. One can take $|\psi\rangle_H$ to be a Laughlin-type wave function describing the ground state of a fractional quantum Hall phase, or the ground state of lattice models such as Kitaev’s honeycomb model, which support protected edge modes. In \cite{Sheffer:2025zyr}, the authors explored this idea to compute the chiral central charge for such theories. 
It would be valuable to place those results on a firmer footing within the general framework of Conjecture~\eqref{eq:conj}. In \cite{Fliss:2017wop}, the entanglement entropy of abelian Chern-Simons theory was computed by extending the Hilbert space to a non-gauge-invariant sector to accommodate conformal edge modes, and by identifying the ground state as a gauge-invariant Ishibashi state in this enlarged Hilbert space. It would be interesting to see whether similar tools can be used to compute multipartite entanglement entropies in Chern-Simons theory, with an extended Hilbert space at the boundaries. 
To make progress in this direction, we need to understand the state of the edge modes at the corner where three regions meet in the ground state.   

Conjecture \eqref{eq:conj} also offers a way to probe the notion of locality more deeply. How essential is locality of the Hamiltonian for the applicability of Conjecture \eqref{eq:conj}? \emph{Any} quantum state $|\psi\rangle_H$ can be engineered as the ground state of \emph{some} Hamiltonian, though not necessarily a local one. Presumably, a generic $q$-partite state constructed in this way will not satisfy Conjecture \eqref{eq:conj}. Conjecture~\eqref{eq:conj} then suggests a more precise, bottom-up definition of a ``topological phase'': namely, a state $|\psi\rangle_H$ that satisfies Conjecture~\eqref{eq:conj}. In this definition, the Hamiltonian - and in particular its locality - plays no role. 
In fact, there are already hints \cite{Akella:2026xza} that locality of the Hamiltonian may be optional for realizing a topological phase. In \cite{Akella:2026xza}, the authors extract the genus of a two-dimensional surface $M$ from the multipartite entanglement of a generic tripartite stabilizer state. 
Of course, the stabilizer generators of a generic stabilizer state need not be local. It would be interesting to pursue this idea further, in particular for higher-partite stabilizer states. As there is no notion of ``size of the region'' for non-local states, enforcing of Conjecture \eqref{eq:conj} is only meaningful without the limit operation on the left-hand side. 
In an abstract characterization of states $|\psi\rangle_H$ satisfying Conjecture~\eqref{eq:conj} (with vanishing correction), Pachner-type moves - implemented as dipole cancellations and additions - relating different GEMs of a given topological manifold should play an important role.

We conclude by listing several research directions aimed at extending the scope of Conjecture~\eqref{eq:conj}.
\begin{itemize}
    \item The Conjecture~\eqref{eq:conj} allows one to extract the absolute value of the TQFT partition function $Z(M)$ for general gapped UV lattice models. For the particular case of Levin-Wen models, which, as was pointed out in Section \ref{sec:extraction}, happen to be free from tri-partite entanglement, we have shown in Appendix \ref{app:alt} that a slightly different approach allows one to extract the full complex value of the partition function. It would be interesting to find a generalization that would work for general lattice models or prove a no-go theorem that would state that it is not possible to extract the argument of $Z(M)$ for arbitrary\footnote{For special \textit{locally-achiral} manifolds it is possible via the approach of \cite{Sheffer:2026dgj}. In the $d=3$ setting, for such manifolds the non-universal UV factors that appear in the multi-invariant $\mathcal{Z}(\sigma_1,\ldots,\sigma_4;|\psi\rangle_H)$ are manifestly real positive and thus do not contribute to the argument. Note that for general $M$ only the tri-partite entanglement part (corresponding to 3-colored subgraphs) of the UV contributions can be non-real-positive. As such contributions are absent in the Levin-Wen case, this does not pose a problem there. } $M$ in such a way.

    \item The natural geometric configuration of $(d+1)$ regions in a $(d-1)$-dimensional space is such that the $(d+1)$ regions do not share a common point. As a result, the genuine $(d+1)$-partite entanglement is necessarily nonlocal. What about the converse? Is all nonlocal entanglement necessarily $(d+1)$-partite, or can it also contain components that are not genuinely $(d+1)$-partite? 
    \item Conjecture~\eqref{eq:conj} relates the genuine $(d+1)$-partite entanglement entropy of the ground state to the partition function of the $d$-dimensional TQFT describing the low-energy physics. One can also divide space into $q>d+1$ regions and compute the genuine $q$-partite entanglement of the state. What aspects of the TQFT are captured by these higher-partite entanglement measures? 
    \item Conjecture~\eqref{eq:conj} concerns multi-invariants that are geometric. However, for a given ground state $|\psi\rangle_H$, one can compute any multi-invariant and its associated signal. What aspects of the TQFT, if any, do these more general quantities capture? In many cases, a TQFT admits extensions to higher-codimension loci, such as boundaries, corners (boundaries of boundaries), and more generally topological defects. It would be interesting to explore whether multipartite entanglement can also characterize this extended structure and make contact with the cobordism hypothesis \cite{Baez:1995xq,Lurie:2009keu}. It is tempting to speculate that non-geometric multi-invariants may be related to such extended TQFT data as well as to decoration of the TQFTs with topological defects and more complicated observables. We plan to return on this question in future work.
    \item If the microscopic theory contains fermions then the low energy theory may be described by spin-TQFT \cite{Chen:2010zpc, Fidkowski_2011, Gu:2012ib, Gu:2013gma, Gaiotto:2015zta}. In addition to the topological class of the manifold, spin-TQFT observables also depend on the spin structure of the manifold. Conjecture \eqref{eq:conj} admits a natural generalization to this case. The right-hand side of equation \eqref{eq:conj} is then not completely independent of the bi-partite triangulation but depends on the discrete spin structure data encoded by the triangulation. It would be interesting to explore this direction further.  
\end{itemize}

\section*{Acknowledgments}
We would like to thank Sriram Akella, Bartek Czech, Francesco Costantino, Kedar Damle, Andrea Grigoletto, Shiraz Minwalla, Onkar Parrikar, Rajath Radhakrishnan, 
Iordanis Romaidis, Yarden Sheffer, Constantin Teleman and Juven Wang  for discussions on related topics. The work of AG is supported by the Department of Atomic Energy, Government of India, under Project Identification No. RTI 4002, and the Infosys Endowment for the study of the Quantum Structure of Spacetime. The work of MDZ is supported by the European Research Council (ERC) under the European Union’s Horizon Europe research and innovation program (grant agreement No. 101171852) and by the VR project grant No. 2023-05590. MDZ also acknowledges the VR Centre for Geometry and Physics (VR grant No. 2022-06593) and the Simons Foundation International Grant for the Simons Collaboration on Global Categorical Symmetries.

\appendix

\section{Fusion properties}
\label{app:fusion}

\subsection{General case}
Throughout this section we consider $n$ to be a fixed positive number. Consider
\begin{equation}
    C^{\hat{\ell}}_L:=\sum_{\ell\in \mathrm{Irr}(\mathcal{C})^{L}}N_{\ell_1\ell_2\ldots \ell_{L}}^{\hat\ell}
    \prod_{i=1}^{L}d_{\ell_i}^{n}
\end{equation}
which is the left-hand side of (\ref{fusion-sum-limit}). We will treat $C_{L}$ as an $L$-dependent vector with components labeled by the simple objects of the fusion category. Next, let us define the matrix $M$ with components
\begin{equation}
    M^j_{\;k}:=\sum_{\ell}N_{k\ell}^j\,d_\ell^n.
\end{equation}
The vectors $C_L$ then satisfy the following recursion relation:
\begin{equation}
    C_{L+1}=MC_L.
\end{equation}
We thus have:
\begin{equation}
    C_{L}=M^LC_0
\end{equation}
where $C^\ell_0=\delta_{\ell\mathbf{1}}$. The matrix $M$ is positive, because for any $j,k$ there exist $\ell$ such that $N_{k\ell}^j>0$. For unitary spherical categories it is known that the unique maximal eigenvalue of the matrix $N^{\bullet}_{\bullet\ell}$, that is the Frobenius-Perron dimension of the object $\ell$, coincides with the quantum dimension $d_\ell$. The corresponding left and right eigenvectors are both multiples of $d_\bullet$, as is clear from the standard relations
\begin{equation}
d_kd_\ell=\sum_{j}N_{k\ell}^jd_j=\sum_{j}N_{j^*\ell}^{k^*}d_j,\qquad d_{j}=d_{j^*}.
\end{equation}
It follows that the maximal eigenvalue of $M$ is 
\begin{equation}
    \lambda=\sum_{\ell}d_\ell^{n+1}
\end{equation}
with the corresponding left and right eigenvectors given again by a multiple of $d_\bullet$.

From the Frobenius-Perron theorem it is known that
\begin{equation}
    \frac{M^L}{\lambda^L}=vw^T+O(e^{-c_nL}),\qquad L\rightarrow\infty,
\end{equation}
where $v$ and $w$ are right and left eigenvectors respectively, normalized such that $w^Tv=1$. The positive constant $c_n$ is the logarithm of the ratio of $\lambda$ over the next-to-maximal eigenvalue of $M$. Thus
\begin{equation}
   (M^L)_{\;k}^j=\lambda^L\frac{d_jd_k}{\mathcal{D}^2}\left(1+O(e^{-c_nL})\right)
\end{equation}
from which (\ref{fusion-sum-limit}) follows.

\subsection{Modular tensor category}
Let us assume that $\mathcal{C}$ is a unitary modular tensor category\footnote{In this case the Turaev-Viro TQFT constructed from $\mathcal{C}$ is equivalent to Reshetikhin-Turaev TQFT constructed from $\mathcal{C}$ stacked with the TQFT complex conjugate to it.}, that is a unitary fusion category with non-degenerate braiding. In this case one can use the Verlinde formula for the fusion coefficients:
\begin{equation}
     N_{\ell_1\ell_2\ldots \ell_{L}}^{\hat\ell}
    =\sum_{r}\frac{S_{\hat{\ell}^*r}\prod_{i=1}^LS_{{\ell}_ir}}{S_{\mathbf{1}r}^{L-1}}
    \label{fusion-verlinde}
\end{equation}
where $S$ is the $S$-matrix, related to the quantum dimensions as
\begin{equation}
    d_\ell=\frac{S_{\ell\mathbf{1}}}{S_{\mathbf{1}\mathbf{1}}},\qquad \mathcal{D}^{-1}=S_{\mathbf{1}\mathbf{1}},
\end{equation}
where $\mathbf{1}$ is the unit object. Plugging (\ref{fusion-verlinde}) into the left-hand side of the conjectural asymptotic formula (\ref{fusion-sum-limit}) we have
\begin{multline}
    \sum_{\ell\in \mathrm{Irr}(\mathcal{C})^{L}}N_{\ell_1\ell_2\ldots \ell_{L}}^{\hat\ell}
    \prod_{i=1}^{L}d_{\ell_i}^{n}=\sum_{r}S_{\hat{\ell}^*r}S_{\mathbf{1}r}\left(\sum_{\ell\in \mathrm{Irr}(\mathcal{C})}d_{\ell}^n\frac{S_{\ell r}}{S_{\mathbf{1}r}}\right)^L=\\
    =\frac{d_\ell}{\mathcal{D}^2}\left(\sum_{\ell\in \mathrm{Irr}(\mathcal{C})}d_{\ell}^{n+1}\right)^L+
    \sum_{r\neq \mathbf{1}}S_{\hat{\ell}^*r}S_{\mathbf{1}r}\left(\sum_{\ell\in \mathrm{Irr}(\mathcal{C})}d_{\ell}^n\frac{S_{\ell r}}{S_{\mathbf{1}r}}\right)^L.
\end{multline}
To prove (\ref{fusion-sum-limit}) it is therefore sufficient to show that all the terms in the sum with $r\neq \mathbf{1}$ are exponentially suppressed relative to the term with $r=\mathbf{1}$ as $L\rightarrow \infty$. From unitarity $d_\ell >0$  and therefore 
\begin{equation}
    \left|\sum_{\ell\in \mathrm{Irr}(\mathcal{C})}d_{\ell}^n\frac{S_{\ell r}}{S_{\mathbf{1}r}}   \right|\leq
    \sum_{\ell\in \mathrm{Irr}(\mathcal{C})}\frac{d_{\ell}^n\,\mathcal{D}}{d_r}|S_{\ell r}|
    \label{ineq-S-sum}
\end{equation}
with the equality only satisfied if, for a fixed $r$, $S_{\ell r}$ has the same phase for all $\ell$. The $S$-matrix elements can be expressed through the topological spins $\theta_\ell$ (the diagonal elements of the $T$-matrix) as follows:
\begin{equation}
    S_{\ell r}=\frac{1}{\mathcal{D}}\sum_{k}N_{\ell^* r}^k\frac{\theta_k}{\theta_\ell \theta_r}\,d_k
\end{equation}
and thus 
\begin{equation}
    |S_{\ell r}|\leq\frac{1}{\mathcal{D}}\sum_{k}N_{\ell^* r}^k\,d_k=\frac{d_\ell d_r}{\mathcal{D}}.
    \label{ineq-S-elem}
\end{equation}
It remains to show that, for any given $r\neq \mathbf{1}$, either inequality (\ref{ineq-S-sum}) is not saturated, or there exists $\ell$ such that (\ref{ineq-S-elem}) is not saturated. Suppose neither of these is true. Then there must exist $r\neq \mathbf{1}$ such that
\begin{equation}
    S_{\ell r}=e^{i\varphi}\frac{d_\ell d_r}{\mathcal{D}}=e^{i\varphi}d_r\,S_{\ell\mathbf{1}}
\end{equation}
for some $\ell$-independent phase $\varphi$. It follows that the $S$-matrix has two linearly dependent columns, which is in contradiction with the assumption that $S$ is invertible, required by the modular tensor category structure.

In this case we have an explicit formula for the ``gap'' constant
\begin{equation}
    c_n=\log\frac{\sum_{\ell\in \mathrm{Irr}(\mathcal{C})}d_{\ell}^{n+1}}{\max\limits_{r\neq \mathbf{1}}\left|\sum_{\ell\in \mathrm{Irr}(\mathcal{C})}d_{\ell}^n\frac{S_{\ell r}}{S_{\mathbf{1}r}}\right|}.
\end{equation}

\section{Alternative approach to extracting the TQFT partition function}
\label{app:alt}

In this appendix we present an alternative method of extracting the TQFT partition function from \textit{normalized} multi-invariants. Consider first (\ref{TV-ent-graph}) for the particular choice of the graph $\Gamma=\Gamma_{S^3}$ consisting of just two vertices (see Figure \ref{fig:3sphere}). It realizes a 3-sphere as a pair of tetrahedra with opposite orientation, glued among the respective faces: $S^3=T_\text{IR}\cup \overline{T_\text{IR}}$. We thus have
\begin{equation}
Z(\Gamma_{S^3};|\psi\rangle)=\left(1+O(e^{-c\,L_\mathrm{min}})\right)\,\mathcal{D}^{-6}\prod_{e\in \mathrm{Edges}(T_\text{IR})}\mathcal{D}^{2L(e)}.
\end{equation}
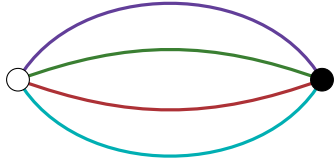
\begin{figure}[ht!]
    \centering
    \begin{tikzpicture}
        \draw[OliveGreen,very thick] (-2,2) to[bend left=20] (2,2);
             \draw[RoyalPurple,very thick] (-2,2) to[bend left=60] (2,2);
              \draw[Maroon,very thick] (-2,2) to[bend right=20] (2,2);
             \draw[TealBlue,very thick] (-2,2) to[bend right=60] (2,2);
                \draw[fill=white] (-2,2) circle (0.15);
                \draw[fill=black] (2,2) circle (0.15);
    \end{tikzpicture}
    \caption{The graph $\Gamma_{S^3}$.}
    \label{fig:3sphere}
\end{figure}
The normalized multi-invariant then reads
\begin{equation}
    \hat{\mathcal{Z}}(\Gamma;|\psi\rangle)\equiv 
    \frac{ \mathcal{Z}(\Gamma;|\psi\rangle)}{\mathcal{Z}(\Gamma_{S^3};|\psi\rangle)^{N_0(\Gamma)/2}}
\end{equation}
where $N_0(\Gamma)$ is the number of vertices of $\Gamma$. The quantities $N_0(\Gamma),\,N_{2:e,n}(\Gamma),\,N_3(\Gamma)$ obey a universal linear relation due to the fact that the Euler characteristic of any closed 3-manifold is zero, and that the number of faces in a triangulation is twice the number of tetrahedra:
\begin{equation}
    N_3(\Gamma)-\sum_{e,n}N_{2:e,n}(\Gamma)+N_0(\Gamma)=0.
    \label{relation-euler}
\end{equation}

For each edge type $e$ we also have a universal relation that comes from the fact that each vertex of the graph $\Gamma$ belongs to a single bi-colored loop with the pair of colors corresponding to $e$:
\begin{equation}
    N_0(\Gamma)=\sum_{n}2n\,N_{2:e,n},\qquad \forall e \in \mathrm{Edges}(T_\text{IR}).
    \label{relation-bi-loops}
\end{equation}
Thus one can always express $N_{2:e,1}$ through $N_0$ and $N_{2:e,n}$ for $n
\geq 2$. Taking into account that $Z_\text{TV}(S^3)=\mathcal{D}^{-2}$ and using the relations (\ref{relation-euler}) and (\ref{relation-bi-loops}) we obtain
\begin{multline}
    \log\hat{\mathcal{Z}}(\Gamma;|\psi\rangle)=\log
    Z_\text{TV}(M_\text{IR})+N_0(\Gamma)\,\log\mathcal{D}\\
    +\sum_{e,\;n\geq 2}N_{2:e,n}(\Gamma)\,L(e)\left(\log\mathcal{D}_{(n)}-n\,\log\mathcal{D}_{(1)}\right)+O(e^{-c\,L_\mathrm{min}}).
    \label{log-normalized}
\end{multline}

In the formulas below we will introduce collective indices $\mu,\nu,\lambda,\ldots$ implicitly running over the set 
\begin{equation}
    \{0\}\cup \{(e,n)\}_{e,\;n\geq 2}
\end{equation}
where, as above, $e$ runs over the six edges of the tetrahedron $T_\text{IR}$ and $n$ runs over positive integers greater that 1.  
With these conventions (\ref{log-normalized}) can be rewritten more compactly as
\begin{equation}
    \log\hat{\mathcal{Z}}(\Gamma;|\psi\rangle)=\log
    Z_\text{TV}(M_\text{IR})+\sum_{\mu,\nu}N_\mu(\Gamma) \Delta^\mu+O(e^{-c\,L_\mathrm{min}})
    \label{log-norm-rep-n-delta}
\end{equation}
where
\begin{equation}
    N_{(e,n)}(\Gamma)\equiv N_{2:e,n}(\Gamma)
\end{equation}
depend on the graph $\Gamma$ only, while
\begin{equation}
    \Delta^0:=\log\mathcal{D},\qquad \Delta^{(e,n)}:=L(e)\left(\log \mathcal{D}_{(n)}-n\,\log\mathcal{D}_{(1)}\right)
\end{equation}
depend on only the input fusion category $\mathcal{C}$ and the lengths of the edges of the regions in the plane subdivision. 

In order to cancel the $\Delta^\nu$-dependent part of (\ref{log-norm-rep-n-delta}) let us consider some fixed collection of graphs $\Gamma^\mu$ such that they all realize a 3-sphere. We want to satisfy
\begin{equation}
    \log\hat{\mathcal{Z}}(\Gamma;|\psi\rangle)-\sum_{\mu}A_\mu(\Gamma)\log\hat{\mathcal{Z}}(\Gamma^\mu;|\psi\rangle)=\log
    Z_\text{TV}(M_\text{IR})+O(e^{-c\,L_\mathrm{min}})
    \label{alternative-formula}
\end{equation}
with some $\Gamma$-dependent (but $\Delta$-independent) coefficients $A_\mu(\Gamma)$ to be determined. The condition above is equivalent to the following linear equation on $A_\mu(\Gamma)$:
\begin{equation}
    2\sum_{\lambda}A_\lambda\Delta^0-\sum_{\lambda,\mu}A_\lambda(\Gamma)N_\mu(\Gamma^\lambda)\Delta^\mu+\sum_{\mu}N_\mu(\Gamma)\Delta^\mu=0.
\end{equation}
Given that $\Delta^\mu$ are generically independent, we thus want to satisfy:
\begin{equation}
    -2\delta_{\mu}^0\sum_{\lambda}A_\lambda+\sum_{\lambda}A_\lambda(\Gamma)N_\mu(\Gamma^\lambda)=N_\mu(\Gamma).
    \label{A-linear-system}
\end{equation}
We will be looking for a choice of the collection $\Gamma^{\lambda}$ that satisfies the following self-consistent truncation property. First, we want that $N_{(e,n)}(\Gamma^{(e',m)})$ if $n>m$. That is $\Gamma^{(e',m)}$ contains bi-colored loops of length only up to $2m$. Let $n_0$ be the smallest number such that $N_{(e,n)}(\Gamma^{0})$ for $n>n_0$. Consider the square matrix $\hat{N}$  with the components $\hat{N}_\mu^\nu:=N_\mu(\Gamma^\lambda)$ where $\mu,\lambda$ run over a truncated set
\begin{equation}
    \{0\}\cup \{(e,n)\}_{e,\;n_*\geq n\geq 2}
\end{equation}
for some $n_*$. We ask that $\hat{N}$ is invertible for any $n_*\geq n_0$. 

The solution of (\ref{A-linear-system}) for $A_\mu(\Gamma)$ for a given $\Gamma$ is then unique can be found in the following way. Take $n_*\geq n_0$ to be such that $N_{(e,n)}(\Gamma)=0$ for $n> n_*$. Then $A_{(e,n)}(\Gamma)=0$ for $n>n_*$ with the other components determined by the following matrix formula:
\begin{equation}
    A(\Gamma)=N(\Gamma)\widehat{N}^{-1}
\end{equation}
where $A(\Gamma)$ and $N(\Gamma)$ are the row-vectors with components $A_\mu(\Gamma)$ and $N_\mu(\Gamma)$ respectively. Below we present a particular collection of the graphs $\Gamma^\lambda$ that satisfy the above properties, as well as the explicit formula for the coefficients  $A_\mu(\Gamma)$ for such a choice.

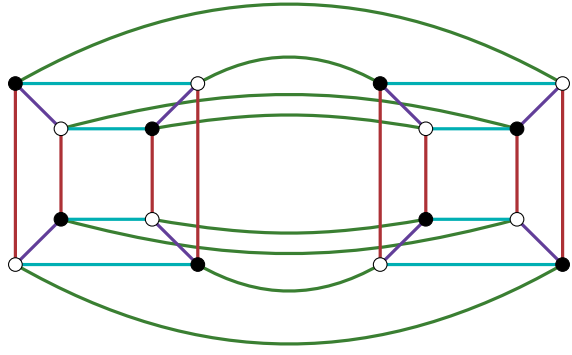
\begin{figure}[ht!]
    \centering
    \begin{tikzpicture}[scale=0.6] 
            \draw[OliveGreen,very thick] (-2,2) to[bend left] (2,2);
            \draw[OliveGreen,very thick] (-6,2) to[bend left] (6,2);
            \draw[OliveGreen,very thick] (-2,-2) to[bend right] (2,-2);
            \draw[OliveGreen,very thick] (-6,-2) to[bend right] (6,-2);

            \draw[OliveGreen,very thick] (-3,1) to[bend left=10] (3,1);
            \draw[OliveGreen,very thick] (-5,1) to[bend left=15] (5,1);
            \draw[OliveGreen,very thick] (-3,-1) to[bend right=10] (3,-1);
            \draw[OliveGreen,very thick] (-5,-1) to[bend right=15] (5,-1);

             \draw[TealBlue,very thick] (-5,1) -- (-3,1);
             \draw[TealBlue,very thick] (-6,2) -- (-2,2);
             \draw[TealBlue,very thick] (-5,-1) -- (-3,-1);
             \draw[TealBlue,very thick] (-6,-2) -- (-2,-2);
             \draw[Maroon,very thick] (-5,-1) -- (-5,1);
            \draw[Maroon,very thick] (-2,-2) -- (-2,2);
            \draw[Maroon,very thick] (-3,-1) -- (-3,1);
            \draw[Maroon,very thick] (-6,-2) -- (-6,2);
             \draw[RoyalPurple,very thick] (-6,-2) -- (-5,-1);
             \draw[RoyalPurple,very thick] (-6,2) -- (-5,1);
             \draw[RoyalPurple,very thick] (-2,-2) -- (-3,-1);
             \draw[RoyalPurple,very thick] (-2,2) -- (-3,1);

             \draw[fill=black] (-3,1) circle (0.15);
             \draw[fill=black] (-2,-2) circle (0.15);
             \draw[fill=black] (-6,2) circle (0.15);
             \draw[fill=black] (-5,-1) circle (0.15);
             \draw[fill=white] (-5,1) circle (0.15);
             \draw[fill=white] (-2,2) circle (0.15);
             \draw[fill=white] (-3,-1) circle (0.15);
             \draw[fill=white] (-6,-2) circle (0.15);

            \draw[TealBlue,very thick] (5,1) -- (3,1);
             \draw[TealBlue,very thick] (6,2) -- (2,2);
             \draw[TealBlue,very thick] (5,-1) -- (3,-1);
             \draw[TealBlue,very thick] (6,-2) -- (2,-2);
             \draw[Maroon,very thick] (5,-1) -- (5,1);
            \draw[Maroon,very thick] (2,-2) -- (2,2);
            \draw[Maroon,very thick] (3,-1) -- (3,1);
            \draw[Maroon,very thick] (6,-2) -- (6,2);
             \draw[RoyalPurple,very thick] (6,-2) -- (5,-1);
             \draw[RoyalPurple,very thick] (6,2) -- (5,1);
             \draw[RoyalPurple,very thick] (2,-2) -- (3,-1);
             \draw[RoyalPurple,very thick] (2,2) -- (3,1);

             \draw[fill=white] (3,1) circle (0.15);
             \draw[fill=white] (2,-2) circle (0.15);
             \draw[fill=white] (6,2) circle (0.15);
             \draw[fill=white] (5,-1) circle (0.15);
             \draw[fill=black] (5,1) circle (0.15);
             \draw[fill=black] (2,2) circle (0.15);
             \draw[fill=black] (3,-1) circle (0.15);
             \draw[fill=black] (6,-2) circle (0.15);

             \end{tikzpicture}
     \caption{The hypercube graph $\Gamma^0$.}
   \label{fig:hypercube}
\end{figure}

As $\Gamma^0$ we take the hypercube graph shown in Figure \ref{fig:hypercube}. As $\Gamma^{(e,n)}$ we take the graphs of the form shown in Figure \ref{fig:graphs-e-n}.
\begin{figure}[ht!]
    \centering
    \begin{subfigure}{0.3\textwidth}
        \centering
        \begin{tikzpicture}             
             \draw[TealBlue,very thick] (45:1) -- (135:1);
             \draw[TealBlue,very thick] (45:-1) -- (135:-1);
             \draw[Maroon,very thick] (-45:1) -- (45:1);
            \draw[Maroon,very thick] (-45:-1) -- (45:-1);
             \draw[TealBlue,very thick] (45:2) -- (135:2);
             \draw[TealBlue,very thick] (45:-2) -- (135:-2);
             \draw[Maroon,very thick] (-45:2) -- (45:2);
            \draw[Maroon,very thick] (-45:-2) -- (45:-2);
             \draw[OliveGreen,very thick] (45:1) to[bend right] (45:2);
             \draw[RoyalPurple,very thick] (45:1) to[bend left] (45:2);

             \draw[OliveGreen,very thick] (-45:1) to[bend right] (-45:2);
             \draw[RoyalPurple,very thick] (-45:1) to[bend left] (-45:2);

             \draw[OliveGreen,very thick] (45:-1) to[bend right] (45:-2);
             \draw[RoyalPurple,very thick] (45:-1) to[bend left] (45:-2);

             \draw[OliveGreen,very thick] (-45:-1) to[bend right] (-45:-2);
             \draw[RoyalPurple,very thick] (-45:-1) to[bend left] (-45:-2);

             \draw[fill=black] (45:1) circle (0.15);
             \draw[fill=black] (45:-1) circle (0.15);
             \draw[fill=black] (-45:2) circle (0.15);
             \draw[fill=black] (-45:-2) circle (0.15);
             \draw[fill=white] (-45:1) circle (0.15);
             \draw[fill=white] (-45:-1) circle (0.15);
             \draw[fill=white] (45:2) circle (0.15);
             \draw[fill=white] (45:-2) circle (0.15);
             
             \end{tikzpicture}
             \caption{$n=2$}
   \end{subfigure}
   \begin{subfigure}{0.3\textwidth}
        \centering
        \begin{tikzpicture}
            \foreach \x in {0,1,...,3} {
             \draw[TealBlue,very thick] (120*\x:1) -- (60+120*\x:1);
             \draw[Maroon,very thick] (60+120*\x:1) -- (120+120*\x:1);

                \draw[TealBlue,very thick] (120*\x:2) -- (60+120*\x:2);
             \draw[Maroon,very thick] (60+120*\x:2) -- (120+120*\x:2);
            }
            
           \foreach \x in {0,1,...,5} {
            \draw[OliveGreen,very thick] (60*\x:1) to[bend right] (60*\x:2);
             \draw[RoyalPurple,very thick] (60*\x:1) to[bend left] (60*\x:2);
            }

              \foreach \x in {0,1,...,2} {
             \draw[fill=black] (120*\x:1) circle (0.15);
             \draw[fill=white] (120*\x:2) circle (0.15);
             \draw[fill=white] (60+120*\x:1) circle (0.15);
             \draw[fill=black] (60+120*\x:2) circle (0.15);
                }

             \end{tikzpicture}
             \caption{$n=3$}
   \end{subfigure}
   \begin{subfigure}{0.3\textwidth}
        \centering
        \begin{tikzpicture}
            \foreach \x in {0,1,...,4} {
             \draw[TealBlue,very thick] (90*\x:1) -- (45+90*\x:1);
             \draw[Maroon,very thick] (45+90*\x:1) -- (90+90*\x:1);

                \draw[TealBlue,very thick] (90*\x:2) -- (45+90*\x:2);
             \draw[Maroon,very thick] (45+90*\x:2) -- (90+90*\x:2);
            }
            
           \foreach \x in {0,1,...,7} {
            \draw[OliveGreen,very thick] (45*\x:1) to[bend right] (45*\x:2);
             \draw[RoyalPurple,very thick] (45*\x:1) to[bend left] (45*\x:2);
            }

              \foreach \x in {0,1,...,3} {
             \draw[fill=black] (90*\x:1) circle (0.15);
             \draw[fill=white] (90*\x:2) circle (0.15);
             \draw[fill=white] (45+90*\x:1) circle (0.15);
             \draw[fill=black] (45+90*\x:2) circle (0.15);
                }

             \end{tikzpicture}
             \caption{$n=4$}
   \end{subfigure}
   \caption{The sequence of graphs $\Gamma^{(e,n)},\;n=2,3,4,\ldots$ for $e$ corresponding to the ${\color{TealBlue}\bullet}{\color{Maroon}\bullet}$ pair of colors.}
   \label{fig:graphs-e-n}
\end{figure}
With this choice of the collection of graphs $\Gamma^\mu$, the explicit solution of the linear system (\ref{A-linear-system}) for $A_\mu(\Gamma)$ reads
\begin{equation}
    A_0(\Gamma)=-\frac{5}{2}N_0(\Gamma)+\sum_{e,\;n\geq 2}\frac{4n-5}{2}N_{2:e,n}(\Gamma),
\end{equation}
\begin{multline}
    A_{(e,2)}(\Gamma)=N_0(\Gamma)+\frac{1}{2}N_{2:e,2}(\Gamma)\\ +\sum_{e',\;n\geq 2}\frac{3n-4}{4}N_{2:e',n}(\Gamma)
    -\sum_{n\geq 2}\frac{n}{4}\left(N_{2:e,n}(\Gamma)+N_{2:e^*,n}(\Gamma)\right),
\end{multline}
\begin{equation}
    A_{(e,n)}(\Gamma)=\frac{1}{2}N_{e,n}(\Gamma),\qquad n\geq 3,
\end{equation}
where $e^*$ denotes the dual edge, i.e. the edge corresponding to the pair of colors complementary to the one for $e$. Note that for any given $\Gamma$ only a finite number of $A_\mu(\Gamma)$ are non-zero. Moreover, using the relations (\ref{relation-bi-loops}) and the total number of bi-colored loops $N_2(\Gamma):=\sum_{e,n}N_{2:e,n}(\Gamma)$ we can rewrite the expressions for $A_0(\Gamma)$ and $A_{(e,2)}$ in the following more compact form:
\begin{equation}
    A_0(\Gamma)=\frac{7}{2}N_0(\Gamma)-\frac{5}{2}N_2(\Gamma)+\frac{1}{2}\sum_{e}N_{2:e,1}(\Gamma),
\end{equation}
\begin{equation}
    A_{(e,2)}(\Gamma)=\frac{1}{2}N_{2:e,2}(\Gamma)-\frac{3}{2}N_0(\Gamma)+N_2(\Gamma)-\frac{1}{4}\sum_{e'\neq e,e^*}N_{2:e',1}(\Gamma).
\end{equation}

\section{``Kaleidoscope''}
\label{app:example}

\begin{figure}[ht!]
        \centering
        \begin{tikzpicture}[scale=0.7]
            \foreach \x in {0,1,...,2} {
             \draw[TealBlue,very thick] (120*\x:0.7) -- (60+120*\x:0.7);
             \draw[Maroon,very thick] (60+120*\x:0.7) -- (120+120*\x:0.7);

                \draw[TealBlue,very thick] (120*\x:2) -- (60+120*\x:2);
             \draw[Maroon,very thick] (60+120*\x:2) -- (120+120*\x:2);
            }
            
           \foreach \x in {0,1,...,5} {
            \draw[OliveGreen,very thick] (60*\x:0.7) to (60*\x:2);
             \draw[RoyalPurple,very thick] (60*\x:0.7) to[bend left=60] (-120+60*\x:2);
            }

              \foreach \x in {1,...,3} {
             \draw[fill=black] (120*\x:0.7) circle (0.15);
             \draw[fill=white] (120*\x:2) circle (0.15);
             \draw[fill=white] (60+120*\x:0.7) circle (0.15);
             \draw (60+120*\x:0.4) node {\tiny \x};
             \draw[fill=black] (60+120*\x:2) circle (0.15);
             \draw (60+120*\x-6:1.7) node {\tiny \x};
                }
                \foreach \x in {4,...,6} {
             \draw (120*\x-6:1.7) node {\tiny \x};
             \draw (120*\x:0.4) node {\tiny \x};
             }
             \end{tikzpicture}
             \quad
              \begin{tikzpicture}[scale=0.7]
            \foreach \x in {0,1,...,2} {
             \draw[TealBlue,very thick] (120*\x:0.7) -- (60+120*\x:0.7);
             \draw[Maroon,very thick] (60+120*\x:0.7) -- (120+120*\x:0.7);

                \draw[TealBlue,very thick] (120*\x:2) -- (60+120*\x:2);
             \draw[Maroon,very thick] (60+120*\x:2) -- (120+120*\x:2);
            }
            
           \foreach \x in {0,1,...,5} {
            \draw[OliveGreen,very thick] (60*\x:0.7) to[bend left] (60*\x:2);
             \draw[RoyalPurple,very thick] (60*\x:0.7) to[bend right] (60*\x:2);
            }

              \foreach \x in {0,1,...,2} {
             \draw[fill=black] (120*\x:0.7) circle (0.15);
             \draw[fill=white] (120*\x:2) circle (0.15);
             \draw[fill=white] (60+120*\x:0.7) circle (0.15);
             \draw[fill=black] (60+120*\x:2) circle (0.15);
                }          
             \end{tikzpicture}
             \quad
              \begin{tikzpicture}[scale=0.7]
            \foreach \x in {0,1,...,2} {
                          \draw[Maroon,very thick] (60+120*\x:0.7) -- (120+120*\x:0.7);

             \draw[Maroon,very thick] (60+120*\x:2) -- (120+120*\x:2);
            }
            
           \foreach \x in {0,1,...,5} {
            \draw[OliveGreen,very thick] (60*\x:0.7) to[bend left] (60*\x:2);
            \draw[TealBlue,very thick] (60*\x:0.7) to[bend right] (60*\x:2);
             \draw[RoyalPurple,very thick] (60*\x:0.7) to[bend left=60] (-120+60*\x:2);
            }

              \foreach \x in {0,1,...,2} {
             \draw[fill=black] (120*\x:0.7) circle (0.15);
             \draw[fill=white] (120*\x:2) circle (0.15);
             \draw[fill=white] (60+120*\x:0.7) circle (0.15);
             \draw[fill=black] (60+120*\x:2) circle (0.15);
                }          
             \end{tikzpicture}
             \quad
              \begin{tikzpicture}[scale=0.7]
            \foreach \x in {0,1,...,2} {
              \draw[Maroon,very thick] (60+120*\x:0.7) -- (120+120*\x:0.7);

             \draw[Maroon,very thick] (60+120*\x:2) -- (120+120*\x:2);
            }
            
           \foreach \x in {0,1,...,5} {
            \draw[OliveGreen,very thick] (60*\x:0.7) to (60*\x:2);
            \draw[TealBlue,very thick] (60*\x:0.7) to[bend left] (60*\x:2);
            \draw[RoyalPurple,very thick] (60*\x:0.7) to[bend right] (60*\x:2);
             
            }

              \foreach \x in {0,1,...,2} {
             \draw[fill=black] (120*\x:0.7) circle (0.15);
             \draw[fill=white] (120*\x:2) circle (0.15);
             \draw[fill=white] (60+120*\x:0.7) circle (0.15);
             \draw[fill=black] (60+120*\x:2) circle (0.15);
                }          
             \end{tikzpicture}

             \vspace{5ex}

             \begin{tikzpicture}[scale=0.7]
            \foreach \x in {0,1,...,2} {
             \draw[TealBlue,very thick] (120*\x:0.7) -- (60+120*\x:0.7);
             
                \draw[TealBlue,very thick] (120*\x:2) -- (60+120*\x:2);
             }
            
           \foreach \x in {0,1,...,5} {
            \draw[OliveGreen,very thick] (60*\x:0.7) to[bend left] (60*\x:2);
            \draw[Maroon,very thick] (60*\x:0.7) to[bend right] (60*\x:2);
            \draw[RoyalPurple,very thick] (60*\x:0.7) to[bend left=60] (-120+60*\x:2);
            }

              \foreach \x in {0,1,...,2} {
             \draw[fill=black] (120*\x:0.7) circle (0.15);
             \draw[fill=white] (120*\x:2) circle (0.15);
             \draw[fill=white] (60+120*\x:0.7) circle (0.15);
             \draw[fill=black] (60+120*\x:2) circle (0.15);
                }          
             \end{tikzpicture}
             \quad
              \begin{tikzpicture}[scale=0.7]
            \foreach \x in {0,1,...,2} {
             \draw[TealBlue,very thick] (120*\x:0.7) -- (60+120*\x:0.7);
           
                \draw[TealBlue,very thick] (120*\x:2) -- (60+120*\x:2);
             }
            
           \foreach \x in {0,1,...,5} {
            \draw[OliveGreen,very thick] (60*\x:0.7) to (60*\x:2);
             \draw[RoyalPurple,very thick] (60*\x:0.7) to[bend left] (60*\x:2);
             \draw[Maroon,very thick] (60*\x:0.7) to[bend right] (60*\x:2);
            }

              \foreach \x in {0,1,...,2} {
             \draw[fill=black] (120*\x:0.7) circle (0.15);
             \draw[fill=white] (120*\x:2) circle (0.15);
             \draw[fill=white] (60+120*\x:0.7) circle (0.15);
             \draw[fill=black] (60+120*\x:2) circle (0.15);
                }          
             \end{tikzpicture}
             \quad
              \begin{tikzpicture}[scale=0.7]

           \foreach \x in {0,1,...,5} {
            \draw[OliveGreen,very thick] (60*\x:0.7) to (60*\x:2);
            \draw[Maroon,very thick] (60*\x:0.7) to[bend left] (60*\x:2);
            \draw[TealBlue,very thick] (60*\x:0.7) to[bend right] (60*\x:2);
             \draw[RoyalPurple,very thick] (60*\x:0.7) to[bend left=60] (-120+60*\x:2);
            }

              \foreach \x in {0,1,...,2} {
             \draw[fill=black] (120*\x:0.7) circle (0.15);
             \draw[fill=white] (120*\x:2) circle (0.15);
             \draw[fill=white] (60+120*\x:0.7) circle (0.15);
             \draw[fill=black] (60+120*\x:2) circle (0.15);
                }          
             \end{tikzpicture}
             \quad
              \begin{tikzpicture}[scale=0.7]

           \foreach \x in {0,1,...,5} {
            \draw[OliveGreen,very thick] (60*\x:0.7) to[bend left=20] (60*\x:2);
             \draw[RoyalPurple,very thick] (60*\x:0.7) to[bend left=60] (60*\x:2);
              \draw[Maroon,very thick] (60*\x:0.7) to[bend right=20] (60*\x:2);
             \draw[TealBlue,very thick] (60*\x:0.7) to[bend right=60] (60*\x:2);
            }
            \foreach \x in {0,1,...,2} {
             \draw[fill=black] (120*\x:0.7) circle (0.15);
             \draw[fill=white] (120*\x:2) circle (0.15);
             \draw[fill=white] (60+120*\x:0.7) circle (0.15);
             \draw[fill=black] (60+120*\x:2) circle (0.15);
                }          
             \end{tikzpicture}

             \vspace{5ex}

             \begin{tikzpicture}[scale=0.7]
            \foreach \x in {0,1,...,2} {
             \draw[TealBlue,very thick] (120*\x:0.7) -- (60+120*\x:0.7);
             \draw[Maroon,very thick] (60+120*\x:0.7) to[bend left] (120+120*\x:0.7);
             \draw[OliveGreen,very thick] (60+120*\x:0.7) to[bend right] (120+120*\x:0.7);

                \draw[TealBlue,very thick] (120*\x:2) -- (60+120*\x:2);
             \draw[Maroon,very thick] (60+120*\x:2) to[bend left] (120+120*\x:2);
             \draw[OliveGreen,very thick] (60+120*\x:2) to[bend right] (120+120*\x:2);
            }
            
           \foreach \x in {0,1,...,5} {
             \draw[RoyalPurple,very thick] (60*\x:0.7) to[bend left=60] (-120+60*\x:2);
            }

              \foreach \x in {0,1,...,2} {
             \draw[fill=black] (120*\x:0.7) circle (0.15);
             \draw[fill=white] (120*\x:2) circle (0.15);
             \draw[fill=white] (60+120*\x:0.7) circle (0.15);
             \draw[fill=black] (60+120*\x:2) circle (0.15);
                }          
             \end{tikzpicture}
             \quad
            \begin{tikzpicture}[scale=0.7]
            \foreach \x in {0,1,...,2} {
             \draw[TealBlue,very thick] (120*\x:0.7) -- (60+120*\x:0.7);
             \draw[Maroon,very thick] (60+120*\x:0.7) to[bend left] (120+120*\x:0.7);
             \draw[OliveGreen,very thick] (60+120*\x:0.7) to[bend right] (120+120*\x:0.7);

                \draw[TealBlue,very thick] (120*\x:2) -- (60+120*\x:2);
             \draw[Maroon,very thick] (60+120*\x:2) to[bend left] (120+120*\x:2);
             \draw[OliveGreen,very thick] (60+120*\x:2) to[bend right] (120+120*\x:2);
            }
            
           \foreach \x in {0,1,...,5} {
             \draw[RoyalPurple,very thick] (60*\x:0.7) to (60*\x:2);
            }

              \foreach \x in {0,1,...,2} {
             \draw[fill=black] (120*\x:0.7) circle (0.15);
             \draw[fill=white] (120*\x:2) circle (0.15);
             \draw[fill=white] (60+120*\x:0.7) circle (0.15);
             \draw[fill=black] (60+120*\x:2) circle (0.15);
                }          
             \end{tikzpicture}
             \quad
              \begin{tikzpicture}[scale=0.7]
            \foreach \x in {0,1,...,2} {
             \draw[Maroon,very thick] (60+120*\x:0.7) to[bend left] (120+120*\x:0.7);
             \draw[OliveGreen,very thick] (60+120*\x:0.7) to[bend right] (120+120*\x:0.7);

             \draw[Maroon,very thick] (60+120*\x:2) to[bend left] (120+120*\x:2);
             \draw[OliveGreen,very thick] (60+120*\x:2) to[bend right] (120+120*\x:2);
            }
            
           \foreach \x in {0,1,...,5} {
             \draw[RoyalPurple,very thick] (60*\x:0.7) to[bend left=60] (-120+60*\x:2);
             \draw[TealBlue,very thick] (60*\x:0.7) to (60*\x:2);
            }

              \foreach \x in {0,1,...,2} {
             \draw[fill=black] (120*\x:0.7) circle (0.15);
             \draw[fill=white] (120*\x:2) circle (0.15);
             \draw[fill=white] (60+120*\x:0.7) circle (0.15);
             \draw[fill=black] (60+120*\x:2) circle (0.15);
                }          
             \end{tikzpicture}
             \quad
              \begin{tikzpicture}[scale=0.7]
            \foreach \x in {0,1,...,2} {
             \draw[Maroon,very thick] (60+120*\x:0.7) to[bend left] (120+120*\x:0.7);
             \draw[OliveGreen,very thick] (60+120*\x:0.7) to[bend right] (120+120*\x:0.7);

             \draw[Maroon,very thick] (60+120*\x:2) to[bend left] (120+120*\x:2);
             \draw[OliveGreen,very thick] (60+120*\x:2) to[bend right] (120+120*\x:2);
            }
            
           \foreach \x in {0,1,...,5} {
             \draw[RoyalPurple,very thick] (60*\x:0.7) to[bend left] (60*\x:2);
             \draw[TealBlue,very thick] (60*\x:0.7) to[bend right] (60*\x:2);
            }

              \foreach \x in {0,1,...,2} {
             \draw[fill=black] (120*\x:0.7) circle (0.15);
             \draw[fill=white] (120*\x:2) circle (0.15);
             \draw[fill=white] (60+120*\x:0.7) circle (0.15);
             \draw[fill=black] (60+120*\x:2) circle (0.15);
                }          
             \end{tikzpicture}

             \vspace{5ex}

 \begin{tikzpicture}[scale=0.7]
            \foreach \x in {0,1,...,2} {
             \draw[TealBlue,very thick] (120*\x:0.7) -- (60+120*\x:0.7);
             \draw[OliveGreen,very thick] (60+120*\x:0.7) -- (120+120*\x:0.7);

                \draw[TealBlue,very thick] (120*\x:2) -- (60+120*\x:2);
             \draw[OliveGreen,very thick] (60+120*\x:2) -- (120+120*\x:2);
            }
            
           \foreach \x in {0,1,...,5} {
            \draw[Maroon,very thick] (60*\x:0.7) to (60*\x:2);
             \draw[RoyalPurple,very thick] (60*\x:0.7) to[bend left=60] (-120+60*\x:2);
            }

              \foreach \x in {0,1,...,2} {
             \draw[fill=black] (120*\x:0.7) circle (0.15);
             \draw[fill=white] (120*\x:2) circle (0.15);
             \draw[fill=white] (60+120*\x:0.7) circle (0.15);
             \draw[fill=black] (60+120*\x:2) circle (0.15);
                }          
             \end{tikzpicture}
             \quad
              \begin{tikzpicture}[scale=0.7]
            \foreach \x in {0,1,...,2} {
             \draw[TealBlue,very thick] (120*\x:0.7) -- (60+120*\x:0.7);
             \draw[OliveGreen,very thick] (60+120*\x:0.7) -- (120+120*\x:0.7);

                \draw[TealBlue,very thick] (120*\x:2) -- (60+120*\x:2);
             \draw[OliveGreen,very thick] (60+120*\x:2) -- (120+120*\x:2);
            }
            
           \foreach \x in {0,1,...,5} {
            \draw[Maroon,very thick] (60*\x:0.7) to[bend left] (60*\x:2);
             \draw[RoyalPurple,very thick] (60*\x:0.7) to[bend right] (60*\x:2);
            }

              \foreach \x in {0,1,...,2} {
             \draw[fill=black] (120*\x:0.7) circle (0.15);
             \draw[fill=white] (120*\x:2) circle (0.15);
             \draw[fill=white] (60+120*\x:0.7) circle (0.15);
             \draw[fill=black] (60+120*\x:2) circle (0.15);
                }          
             \end{tikzpicture}
             \quad
              \begin{tikzpicture}[scale=0.7]
            \foreach \x in {0,1,...,2} {
                          \draw[OliveGreen,very thick] (60+120*\x:0.7) -- (120+120*\x:0.7);

             \draw[OliveGreen,very thick] (60+120*\x:2) -- (120+120*\x:2);
            }
            
           \foreach \x in {0,1,...,5} {
            \draw[Maroon,very thick] (60*\x:0.7) to[bend left] (60*\x:2);
            \draw[TealBlue,very thick] (60*\x:0.7) to[bend right] (60*\x:2);
             \draw[RoyalPurple,very thick] (60*\x:0.7) to[bend left=60] (-120+60*\x:2);
            }

              \foreach \x in {0,1,...,2} {
             \draw[fill=black] (120*\x:0.7) circle (0.15);
             \draw[fill=white] (120*\x:2) circle (0.15);
             \draw[fill=white] (60+120*\x:0.7) circle (0.15);
             \draw[fill=black] (60+120*\x:2) circle (0.15);
                }          
             \end{tikzpicture}
             \quad
              \begin{tikzpicture}[scale=0.7]
            \foreach \x in {0,1,...,2} {
              \draw[OliveGreen,very thick] (60+120*\x:0.7) -- (120+120*\x:0.7);

             \draw[OliveGreen,very thick] (60+120*\x:2) -- (120+120*\x:2);
            }
            
           \foreach \x in {0,1,...,5} {
            \draw[Maroon,very thick] (60*\x:0.7) to (60*\x:2);
            \draw[TealBlue,very thick] (60*\x:0.7) to[bend left] (60*\x:2);
            \draw[RoyalPurple,very thick] (60*\x:0.7) to[bend right] (60*\x:2);
             
            }

              \foreach \x in {0,1,...,2} {
             \draw[fill=black] (120*\x:0.7) circle (0.15);
             \draw[fill=white] (120*\x:2) circle (0.15);
             \draw[fill=white] (60+120*\x:0.7) circle (0.15);
             \draw[fill=black] (60+120*\x:2) circle (0.15);
                }          
             \end{tikzpicture}

   \caption{The multi-invariants corresponding to the 16 terms in the decomposition (\ref{4-signal-explicit}), ordered in the same way,  in the case when the graph $\Gamma$ (the first one in the list), corresponding to the quadruple $(\sigma_1,\sigma_2,\sigma_3,\sigma_4)$, is realizing the Lens space $L(3,1)$.  Explicitly, assuming the labeling of the vertices shown in the figure, the permutations are given by $\sigma_1=(1)(2)(3)(4)(5)(6)$, $\sigma_2=(15)(26)(34)$, $\sigma_3=(14)(25)(36)$, $\sigma_4=(132)(465)$.}
   \label{fig:L31-kaleidoscope}
\end{figure}
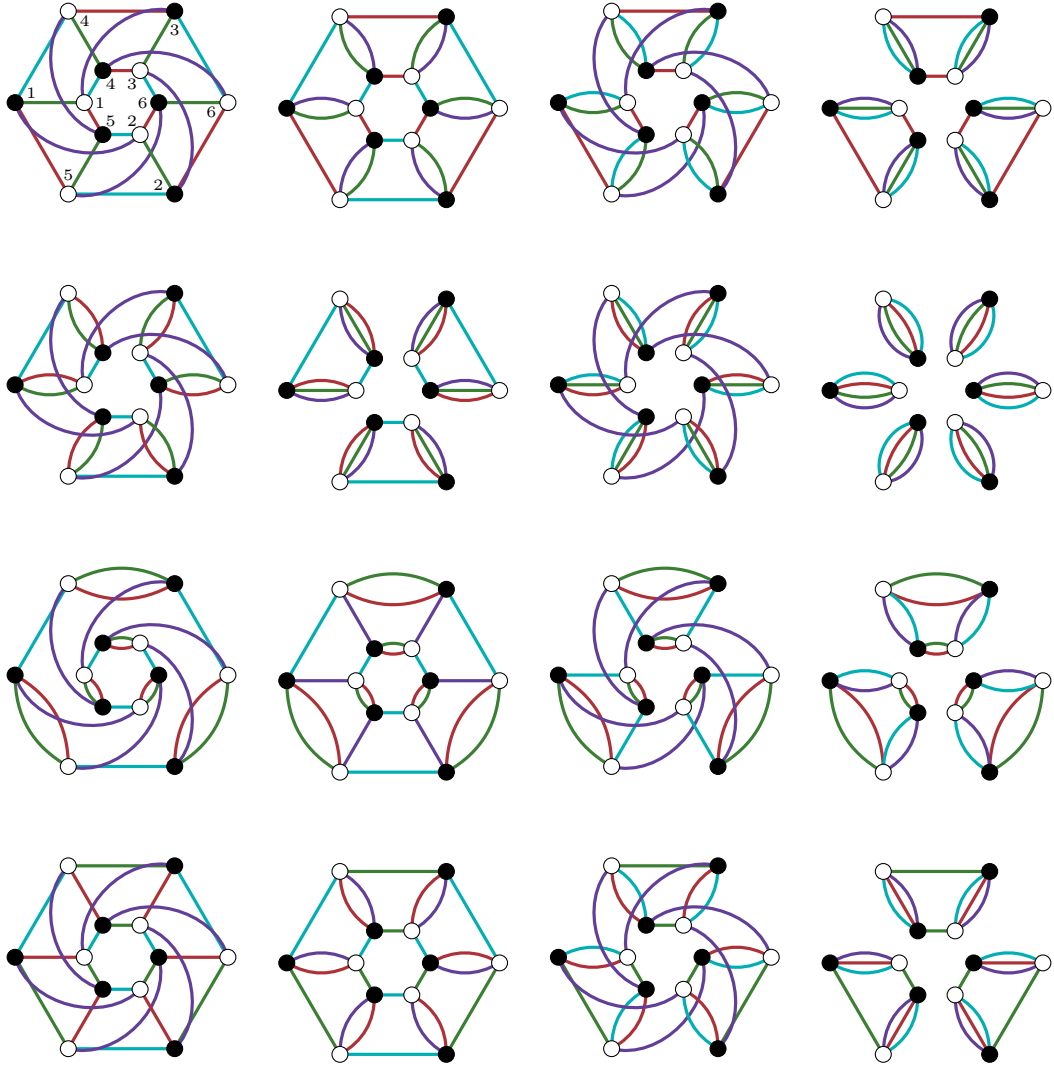

The 4-partite signal (\ref{genuine-4-specific}) times two, by definition, explicitly reads:
\begin{multline}
 \mathcal{E}(\sigma_1-\sigma_2, \sigma_2-\sigma_1,\sigma_3-\sigma_1,\sigma_4-\sigma_1;\cdot)=
 \\
  \mathcal{E}(\sigma_1, \sigma_2,\sigma_3,\sigma_4;\cdot) 
  -
   \mathcal{E}(\sigma_1, \sigma_2,\sigma_3,\sigma_1;\cdot) 
   -
    \mathcal{E}(\sigma_1, \sigma_2,\sigma_1,\sigma_4;\cdot) 
    +
     \mathcal{E}(\sigma_1, \sigma_2,\sigma_1,\sigma_1;\cdot) 
     \\
     -\mathcal{E}(\sigma_1, \sigma_1,\sigma_3,\sigma_4;\cdot) 
     +
     \mathcal{E}(\sigma_1, \sigma_1,\sigma_3,\sigma_1;\cdot) 
     +
     \mathcal{E}(\sigma_1, \sigma_1,\sigma_1,\sigma_4;\cdot) 
     -
     \mathcal{E}(\sigma_1, \sigma_1,\sigma_1,\sigma_1;\cdot) 
     \\
     -\mathcal{E}(\sigma_2, \sigma_2,\sigma_3,\sigma_4;\cdot) 
     +
     \mathcal{E}(\sigma_2, \sigma_2,\sigma_3,\sigma_1;\cdot) 
     +
     \mathcal{E}(\sigma_2, \sigma_2,\sigma_1,\sigma_4;\cdot) 
     -
     \mathcal{E}(\sigma_2, \sigma_2,\sigma_1,\sigma_1;\cdot) 
     \\
     +
     \mathcal{E}(\sigma_2, \sigma_1,\sigma_3,\sigma_4;\cdot) 
     -
     \mathcal{E}(\sigma_2, \sigma_1,\sigma_3,\sigma_1;\cdot) 
     -
     \mathcal{E}(\sigma_2, \sigma_1,\sigma_1,\sigma_4;\cdot) 
     +
     \mathcal{E}(\sigma_2, \sigma_1,\sigma_1,\sigma_1;\cdot).
  \label{4-signal-explicit}
\end{multline}

If one takes the quadruple $(\sigma_1,\sigma_2,\sigma_3,\sigma_4)$ to be corresponding to a graph realizing a Lens space $L(3,1)$ (as the one in e.g. \cite{FerriGagliardi1982}), the 16 terms in the expansion above will correspond to the 16 graphs shown in Figure \ref{fig:L31-kaleidoscope}.

The cancellation of the non-trivial UV factors from (\ref{uv_contribution}) among the 16 terms (after taking the logarithm), arranged in the same way,  works as follows.
\begin{equation}
    L({\color{Maroon}\bullet}{\color{OliveGreen}\bullet})\log \mathcal{D}_{(1)}\times\qquad
    \begin{array}{rccll}
         (+0 & -0& -0 & -0 & \\
         -6 & +6& +6 & -6 & \\
         -6 & +6& +6 & -6 & \\
         +0 & -0& -0 & -0) & =0,\\
    \end{array}
\end{equation}
\begin{equation}
    L({\color{Maroon}\bullet}{\color{OliveGreen}\bullet})\log \mathcal{D}_{(2)}\times\qquad
    \begin{array}{rccll}
         (+3 & -3& -3 & +3 & \\
         -0 & +0& +0 & -0 & \\
         -0 & +0& +0 & -0 & \\
         +3 & -3& -3 & +3) & =0,\\
    \end{array}
\end{equation}
\begin{equation}
    L({\color{RoyalPurple}\bullet}{\color{OliveGreen}\bullet})\log \mathcal{D}_{(1)}\times\qquad
    \begin{array}{rccll}
         (+0 & -6& -0 & +6 & \\
         -0 & +6& +0 & -6 & \\
         -0 & +0& +0 & -0 & \\
         +0 & -0& -0 & +0) & =0,\\
    \end{array}
\end{equation}
\begin{equation}
    L({\color{RoyalPurple}\bullet}{\color{OliveGreen}\bullet})\log \mathcal{D}_{(2)}\times\qquad
    \begin{array}{rccll}
         (+0 & -0& -0 & +0 & \\
         -0 & +0& +0 & -0 & \\
         -2 & +2& +2 & -2 & \\
         +2 & -2& -2 & +2)& =0,\\
    \end{array}
\end{equation}
\begin{equation}
    L({\color{RoyalPurple}\bullet}{\color{OliveGreen}\bullet})\log \mathcal{D}_{(3)}\times\qquad
    \begin{array}{rccll}
         (+3 & -0& -3 & +0 & \\
         -3 & +0& +3 & -0 & \\
         -0 & +0& +0 & -0 & \\
         +0 & -0& -0 & +0) & =0,\\
    \end{array}
\end{equation}
\begin{equation}
    L({\color{RoyalPurple}\bullet}{\color{Maroon}\bullet})\log \mathcal{D}_{(1)}\times\qquad
    \begin{array}{rccll}
         (+0 & -0& -0 & +0 & \\
         -0 & +6& +0 & -6 & \\
         -0 & +0& +0 & -0 & \\
         +0 & -6& -0 & +6) & =0,\\
    \end{array}
\end{equation}
\begin{equation}
    L({\color{RoyalPurple}\bullet}{\color{Maroon}\bullet})\log \mathcal{D}_{(2)}\times\qquad
    \begin{array}{rccll}
         (+3 & -3& -3 & +3 & \\
         -0 & +0& +0 & -0 & \\
         -3 & +3& +3 & -3 & \\
         +0 & -0& -0 & +0) & =0,\\
    \end{array}
\end{equation}
\begin{equation}
    L({\color{RoyalPurple}\bullet}{\color{Maroon}\bullet})\log \mathcal{D}_{(3)}\times\qquad
    \begin{array}{rccll}
         (+0 & -0& -0 & +0 & \\
         -2 & +0& +2 & -0 & \\
         -0 & +0& +0 & -0 & \\
         +2 & -0& -2 & +0) & =0,\\
    \end{array}
\end{equation}
\begin{equation}
    L({\color{TealBlue}\bullet}{\color{Maroon}\bullet})\log \mathcal{D}_{(1)}\times\qquad
    \begin{array}{rccll}
         (+0 & -0& -0 & +0 & \\
         -0 & +0& +6 & -6 & \\
         -0 & +0& +0 & -0 & \\
         +0 & -0& -6 & +6) & =0,\\
    \end{array}
\end{equation}
\begin{equation}
    L({\color{TealBlue}\bullet}{\color{Maroon}\bullet})\log \mathcal{D}_{(2)}\times\qquad
    \begin{array}{rccll}
         (+0 & -0& -3 & +3 & \\
         -3 & +3& +0 & -0 & \\
         -0 & +0& +3 & -3 & \\
         +3 & -3& -0 & +0) & =0,\\
    \end{array}
\end{equation}
\begin{equation}
    L({\color{TealBlue}\bullet}{\color{Maroon}\bullet})\log \mathcal{D}_{(3)}\times\qquad
    \begin{array}{rccll}
         (+2 & -2& -0 & +0 & \\
         -0 & +0& +0 & -0 & \\
         -2 & +2& +0 & -0 & \\
         +0 & -0& -0 & +0) & =0,\\
    \end{array}
\end{equation}
\begin{equation}
    L({\color{TealBlue}\bullet}{\color{RoyalPurple}\bullet})\log \mathcal{D}_{(1)}\times\qquad
    \begin{array}{rccll}
         (+0 & -0& -0 & +6 & \\
         -0 & +0& +0 & -6 & \\
         -0 & +0& +0 & -6 & \\
         +0 & -0& -0 & +6) & =0,\\
    \end{array}
\end{equation}
\begin{equation}
    L({\color{TealBlue}\bullet}{\color{RoyalPurple}\bullet})\log \mathcal{D}_{(2)}\times\qquad
    \begin{array}{rccll}
         (+3 & -3& -0 & +0 & \\
         -3 & +3& +0 & -0 & \\
         -3 & +3& +0 & -0 & \\
         +3 & -3& -0 & +0) & =0,\\
    \end{array}
\end{equation}
\begin{equation}
    L({\color{TealBlue}\bullet}{\color{RoyalPurple}\bullet})\log \mathcal{D}_{(3)}\times\qquad
    \begin{array}{rccll}
         (+0 & -0& -2 & +0 & \\
         -0 & +0& +2 & -0 & \\
         -0 & +0& +2 & -0 & \\
         +0 & -0& -2 & +0) & =0,\\
    \end{array}
\end{equation}
\begin{equation}
    L({\color{TealBlue}\bullet}{\color{OliveGreen}\bullet})\log \mathcal{D}_{(1)}\times\qquad
    \begin{array}{rccll}
         (+0 & -0& -6 & +6 & \\
         -0 & +0& +6 & -6 & \\
         -2 & +2& +0 & -0 & \\
         +2 & -2& -0 & +0) & =0,\\
    \end{array}
\end{equation}
\begin{equation}
    L({\color{TealBlue}\bullet}{\color{OliveGreen}\bullet})\log \mathcal{D}_{(2)}\times\qquad
    \begin{array}{rccll}
         (+3 & -3& -0 & +0 & \\
         -3 & +3& +0 & -0 & \\
         -0 & +0& +3 & -3 & \\
         +0 & -0& -3 & +3) & =0,\\
    \end{array}
\end{equation}
\begin{equation}
    L({\color{TealBlue}\bullet}{\color{OliveGreen}\bullet})\log \mathcal{D}_{(3)}\times\qquad
    \begin{array}{rccll}
         (+0 & -0& -0 & +0 & \\
         -0 & +0& +0 & -0 & \\
         -0 & +0& +0 & -0 & \\
         +0 & -0& -0 & +0) & =0,\\
    \end{array}
\end{equation}
and the obvious
\begin{equation}
    -2\log \mathcal{D}\times\qquad
    \begin{array}{rccll}
         (+12 & -12 & -12 & +12 & \\
         -12 & +12& +12 & -12 & \\
         -12 & +12& +12 & -12 & \\
         +12 & -12& -12 & +12) & =0.\\
    \end{array}
\end{equation}
The IR TQFT partition functions contribute as follows:
\begin{multline}
        \begin{array}{rccll}
         (+\log Z(L(3,1)) & -\log Z(S^3) & -\log Z(S^3) & +3\log Z(S^3) & \\
         -\log Z(S^3) & +3\log Z(S^3)& +2\log Z(S^3) & -6\log Z(S^3) & \\
         -\log Z(S^3) & +\log Z(S^3)& +\log Z(S^3) & -3\log Z(S^3) & \\
         +\log Z(\overline{L(3,1)}) & -\log Z(S^3)& -\log Z(S^3) & +3\log Z(S^3)) & =\\
    \end{array}
    \\
    =\log Z(L(3,1))+\log \overline{Z(L(3,1))}  -2\log Z(S^3).
\end{multline}

Finally we note that if the same is applied to the hypercube graph in Figure~\ref{fig:hypercube} realizing $S^3$, instead of the first graph in Figure~\ref{fig:L31-kaleidoscope} realizing $L(3,1)$, the result, after all the cancellations, is $-2\log Z(S^3)$. This, in particular, allows one to obtain the $Z(S^3)$ value that appears in the general relation (\ref{eq:conj}).

\section{Examples of GEMs}
\label{app:more-examples}

\begin{figure}[ht!]
    \centering
    \begin{tikzpicture}[scale=0.7]

        \coordinate (J) at (0:1);
        \coordinate (k) at (60:1);
        \coordinate (H) at (120:1);
        \coordinate (i) at (180:1);
        \coordinate (A) at (240:1);
        \coordinate (b) at (300:1);

        \coordinate (d) at (0:2);
        \coordinate (F) at (60:2);
        \coordinate (e) at (120:2);
        \coordinate (D) at (180:2);
        \coordinate (f) at (240:2);
        \coordinate (E) at (300:2);

        \coordinate (a) at (0:5);
        \coordinate (I) at (60:5);
        \coordinate (j) at (120:5);
        \coordinate (B) at (180:5);
        \coordinate (h) at (240:5);
        \coordinate (K) at (300:5);

        \coordinate (g) at (70:3.5);
        \coordinate (L) at (110:3.5);
        \coordinate (c) at (190:3.5);
        \coordinate (G) at (230:3.5);
        \coordinate (C) at (-10:3.5);
        \coordinate (l) at (-50:3.5);

        \draw[RoyalPurple,very thick] (a) to[bend left] (A);
        \draw[RoyalPurple,very thick] (b) to[bend left] (B);
        \draw[RoyalPurple,very thick] (c) to[bend left=80] (C);
        \draw[RoyalPurple,very thick] (d) to[bend left] (D);
        \draw[RoyalPurple,very thick] (e) to[bend left] (E);
        \draw[RoyalPurple,very thick] (f) to[bend left] (F);
        \draw[RoyalPurple,very thick] (g) to[bend left=80] (G);
        \draw[RoyalPurple,very thick] (h) to[bend left] (H);
        \draw[RoyalPurple,very thick] (i) to[bend left] (I);
        \draw[RoyalPurple,very thick] (j) to[bend left] (J);
        \draw[RoyalPurple,very thick] (k) to[bend left] (K);
        \draw[RoyalPurple,very thick] (l) to[bend left=80] (L);

        \draw[OliveGreen,very thick] (e) -- (D);
        \draw[OliveGreen,very thick] (i) -- (H);
        \draw[OliveGreen,very thick] (k) -- (J);
        \draw[OliveGreen,very thick] (F) -- (d);
        \draw[OliveGreen,very thick] (A) -- (b);
        \draw[OliveGreen,very thick] (f) -- (E);
        \draw[OliveGreen,very thick] (j) -- (L);
        \draw[OliveGreen,very thick] (g) -- (I);
        \draw[OliveGreen,very thick] (a) -- (C);
        \draw[OliveGreen,very thick] (l) -- (K);
        \draw[OliveGreen,very thick] (h) -- (G);
        \draw[OliveGreen,very thick] (c) -- (B);

        \draw[TealBlue,very thick] (k) -- (H);
        \draw[TealBlue,very thick] (i) -- (A);
        \draw[TealBlue,very thick] (b) -- (J);
        \draw[TealBlue,very thick] (g) -- (F);
        \draw[TealBlue,very thick] (e) -- (L);
        \draw[TealBlue,very thick] (c) -- (D);
        \draw[TealBlue,very thick] (f) -- (G);
        \draw[TealBlue,very thick] (l) -- (E);
        \draw[TealBlue,very thick] (d) -- (C);
        \draw[TealBlue,very thick] (a) -- (I);
        \draw[TealBlue,very thick] (j) -- (B);
        \draw[TealBlue,very thick] (h) -- (K);

        \draw[Maroon,very thick] (k) -- (F);
        \draw[Maroon,very thick] (d) -- (J);
        \draw[Maroon,very thick] (e) -- (H);
        \draw[Maroon,very thick] (i) -- (D);
        \draw[Maroon,very thick] (f) -- (A);
        \draw[Maroon,very thick] (b) -- (E);
        \draw[Maroon,very thick] (L) -- (g);
        \draw[Maroon,very thick] (j) -- (I);
        \draw[Maroon,very thick] (c) -- (G);
        \draw[Maroon,very thick] (h) -- (B);
        \draw[Maroon,very thick] (l) -- (C);
        \draw[Maroon,very thick] (a) -- (K);

        \draw[fill=white] (l) circle (0.15);
        \draw[fill=white] (a) circle (0.15);
        \draw[fill=black] (K) circle (0.15);
        \draw[fill=black] (C) circle (0.15);
        \draw[fill=white] (b) circle (0.15);
        \draw[fill=white] (k) circle (0.15);
        \draw[fill=black] (A) circle (0.15);
        \draw[fill=black] (J) circle (0.15);

        \draw[fill=black] (D) circle (0.15);
        \draw[fill=black] (I) circle (0.15);
        \draw[fill=white] (g) circle (0.15);
        \draw[fill=white] (d) circle (0.15);
        \draw[fill=black] (H) circle (0.15);
        \draw[fill=black] (E) circle (0.15);
        \draw[fill=white] (f) circle (0.15);
        \draw[fill=white] (h) circle (0.15);

        \draw[fill=black] (B) circle (0.15);
        \draw[fill=white] (c) circle (0.15);
        \draw[fill=white] (e) circle (0.15);
        \draw[fill=black] (G) circle (0.15);
        
        \draw[fill=black] (L) circle (0.15);
        \draw[fill=white] (j) circle (0.15);
        \draw[fill=white] (i) circle (0.15);
        \draw[fill=black] (F) circle (0.15);

    \end{tikzpicture}
    \caption{The minimal GEM realizing $M=T^3$. Note that the corresponding TQFT partition function $Z(T^3)$ has the meaning of the ground state degeneracy on $T^2$, as well as the number of anyons (simple line operators).}
    \label{fig:3-torus}
\end{figure}
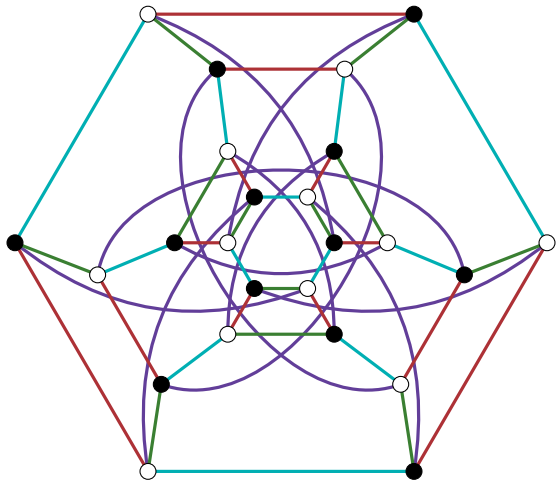

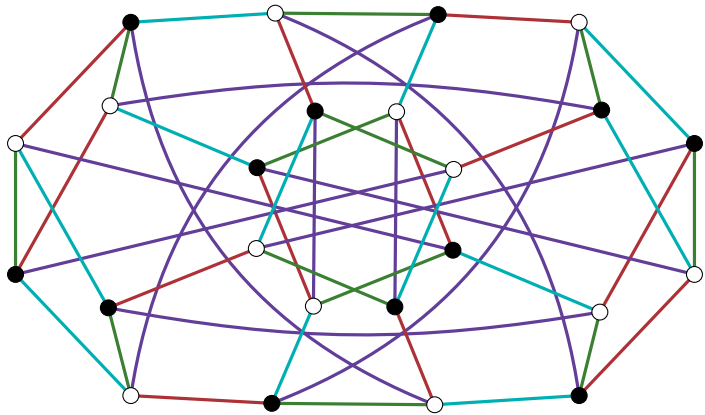
\begin{figure}[ht!]
    \centering
    \begin{tikzpicture}[scale=0.7]

        \coordinate (l) at (22:2);
        \coordinate (a) at (67:2);
        \coordinate (K) at (112:2);
        \coordinate (C) at (157:2);
        \coordinate (b) at (202:2);
        \coordinate (k) at (247:2);
        \coordinate (A) at (292:2);
        \coordinate (J) at (337:2);

        \coordinate (D) at (22:5);
        \coordinate (I) at (67:4);
        \coordinate (g) at (112:4);
        \coordinate (d) at (157:5);
        \coordinate (H) at (202:5);
        \coordinate (E) at (247:4);
        \coordinate (f) at (292:4);
        \coordinate (h) at (337:5);

        \coordinate (B) at (11:6.5);
        \coordinate (c) at (-11:6.5);
        \coordinate (e) at (40:5.5);
        \coordinate (G) at (-40:5.5);
        
        \coordinate (L) at (11:-6.5);
        \coordinate (j) at (-11:-6.5);
        \coordinate (i) at (40:-5.5);
        \coordinate (F) at (-40:-5.5);

        \draw[Maroon,very thick] (L) -- (d);
        \draw[Maroon,very thick] (C) -- (k);
        \draw[Maroon,very thick] (a) -- (J);
        \draw[Maroon,very thick] (B) -- (h);
        \draw[Maroon,very thick] (F) -- (j);
        \draw[Maroon,very thick] (G) -- (c);
        \draw[Maroon,very thick] (I) -- (e);
        \draw[Maroon,very thick] (i) -- (E);

        \draw[RoyalPurple,very thick] (a) -- (A);
        \draw[RoyalPurple,very thick] (b) -- (B);
        \draw[RoyalPurple,very thick] (c) -- (C);
        \draw[RoyalPurple,very thick] (d) to[bend left=10] (D);
        \draw[RoyalPurple,very thick] (e) to[bend left] (E);
        \draw[RoyalPurple,very thick] (f) to[bend left] (F);
        \draw[RoyalPurple,very thick] (g) to[bend left] (G);
        \draw[RoyalPurple,very thick] (h) to[bend left =10] (H);
        \draw[RoyalPurple,very thick] (i) to[bend left] (I);
        \draw[RoyalPurple,very thick] (j) -- (J);
        \draw[RoyalPurple,very thick] (k) -- (K);
        \draw[RoyalPurple,very thick] (l) -- (L);

        \draw[OliveGreen,very thick] (C) -- (a);
        \draw[OliveGreen,very thick] (k) -- (J);
        \draw[OliveGreen,very thick] (g) -- (I);
        \draw[OliveGreen,very thick] (E) -- (f);
        \draw[OliveGreen,very thick] (B) -- (c);
        \draw[OliveGreen,very thick] (h) -- (G);
        \draw[OliveGreen,very thick] (F) -- (d);
        \draw[OliveGreen,very thick] (j) -- (L);
        
        \draw[TealBlue,very thick] (a) -- (I);
        \draw[TealBlue,very thick] (k) -- (E);
        \draw[TealBlue,very thick] (e) -- (B);
        \draw[TealBlue,very thick] (J) -- (h);
        \draw[TealBlue,very thick] (f) -- (G);
        \draw[TealBlue,very thick] (d) -- (C);
        \draw[TealBlue,very thick] (F) -- (g);
        \draw[TealBlue,very thick] (L) -- (i);
        
        \draw[OliveGreen,very thick] (K) -- (l);
        \draw[OliveGreen,very thick] (b) -- (A);
        \draw[OliveGreen,very thick] (e) -- (D);
        \draw[OliveGreen,very thick] (H) -- (i);

        \draw[TealBlue,very thick] (K) -- (b);
        \draw[TealBlue,very thick] (A) -- (l);
        \draw[TealBlue,very thick] (D) -- (c);
        \draw[TealBlue,very thick] (j) -- (H);

        \draw[Maroon,very thick] (l) -- (D);
        \draw[Maroon,very thick] (g) -- (K);
        \draw[Maroon,very thick] (A) -- (f);
        \draw[Maroon,very thick] (H) -- (b);

        \draw[fill=white] (l) circle (0.15);
        \draw[fill=white] (a) circle (0.15);
        \draw[fill=black] (K) circle (0.15);
        \draw[fill=black] (C) circle (0.15);
        \draw[fill=white] (b) circle (0.15);
        \draw[fill=white] (k) circle (0.15);
        \draw[fill=black] (A) circle (0.15);
        \draw[fill=black] (J) circle (0.15);

        \draw[fill=black] (D) circle (0.15);
        \draw[fill=black] (I) circle (0.15);
        \draw[fill=white] (g) circle (0.15);
        \draw[fill=white] (d) circle (0.15);
        \draw[fill=black] (H) circle (0.15);
        \draw[fill=black] (E) circle (0.15);
        \draw[fill=white] (f) circle (0.15);
        \draw[fill=white] (h) circle (0.15);

        \draw[fill=black] (B) circle (0.15);
        \draw[fill=white] (c) circle (0.15);
        \draw[fill=white] (e) circle (0.15);
        \draw[fill=black] (G) circle (0.15);
        
        \draw[fill=black] (L) circle (0.15);
        \draw[fill=white] (j) circle (0.15);
        \draw[fill=white] (i) circle (0.15);
        \draw[fill=black] (F) circle (0.15);

    \end{tikzpicture}
    \caption{The minimal GEM realizing Poincar\'e homology sphere $M=S^3/2I$, the quotient of the sphere by the action of the binary icosahedral group.}
    \label{fig:poincare-sphere}
\end{figure}

Figures \ref{fig:s2s1}, \ref{fig:3-torus}, \ref{fig:poincare-sphere} show the unique minimal GEMs (\textit{superattractors} in the terminology of \cite{Lins1995}) for $S^2\times S^1$, $T^3$ and Poincar\'e homology sphere respectively. The unique minimal GEM realizing $L(3,1)$ Lens space has appeared as the first entry in Figure \ref{fig:L31-kaleidoscope}. The minimal GEM realizing $S^3$ has appeared in Figure \ref{fig:3sphere}, while some of its non-minimal GEMs are shown in Figures \ref{fig:hypercube} and \ref{fig:graphs-e-n}. 

\begin{figure}[ht!]
    \centering
    \begin{tikzpicture}[scale=0.7]

        \coordinate (a) at (2,2);
        \coordinate (b) at (-1,1);
        \coordinate (c) at (-2,-2);
        \coordinate (d) at (1,-1);

        \coordinate (A) at (1,1);
        \coordinate (B) at (-2,2);
        \coordinate (C) at (-1,-1);
        \coordinate (D) at (2,-2);

        \draw[OliveGreen,very thick] (b) -- (C);
        \draw[OliveGreen,very thick] (B) -- (c);
        \draw[OliveGreen,very thick] (a) to[bend left] (A);
        \draw[OliveGreen,very thick] (d) to[bend left] (D);

        \draw[TealBlue,very thick] (c) -- (D);
        \draw[TealBlue,very thick] (C) -- (d);
        \draw[TealBlue,very thick] (A) to[bend left] (a);
        \draw[TealBlue,very thick] (B) to[bend left] (b);

        \draw[Maroon,very thick] (a) -- (D);
        \draw[Maroon,very thick] (A) -- (d);
        \draw[Maroon,very thick] (b) to[bend left] (B);
        \draw[Maroon,very thick] (c) to[bend left] (C);

        \draw[RoyalPurple,very thick] (a) -- (B);
        \draw[RoyalPurple,very thick] (A) -- (b);
        \draw[RoyalPurple,very thick] (D) to[bend left] (d);
        \draw[RoyalPurple,very thick] (C) to[bend left] (c);

        \draw[fill=white] (a) circle (0.15);
        \draw[fill=black] (C) circle (0.15);
        \draw[fill=white] (b) circle (0.15);
        \draw[fill=black] (A) circle (0.15);
        
        \draw[fill=black] (D) circle (0.15);
        \draw[fill=white] (d) circle (0.15);
        
        \draw[fill=black] (B) circle (0.15);
        \draw[fill=white] (c) circle (0.15);

    \end{tikzpicture}
    \caption{The minimal GEM realizing $M=S^2\times S^1$. For any simple TQFT (that is, with the unique vacuum, or equivalently, with no non-trivial local operators the corresponding partition function is one: $Z(S^2\times S^1)=1$.}
    \label{fig:s2s1}
\end{figure}
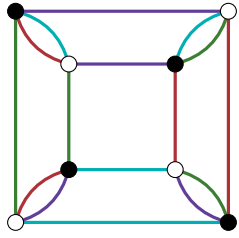

\clearpage

\bibliographystyle{JHEP}
\bibliography{refs.bib}

\end{document}